%% file: ANA-FDET-2019-06-PAPER.tex
\pdfoutput=1
\documentclass[PAPER=true,atlasdraft=false,cernpreprint,txfonts=true,UKenglish, texmf, orcidlogo]{atlasdoc}
\usepackage{atlaspackage}
\usepackage{atlasbiblatex}
\usepackage{atlasjournal}
\graphicspath{{logos/}{figuresSEP/}}
\usepackage{ANA-FDET-2019-06-PAPER-defs}

\AtlasTitle{Performance of the ATLAS forward proton Time-of-Flight detector in Run 2}

\AtlasAbstract{
We present performance studies of the Time-of-Flight (ToF) subdetector of the ATLAS Forward Proton (AFP) detector at the LHC. Efficiencies and resolutions are measured using high-statistics data samples collected at low and moderate pile-up in 2017, the first year when the detectors were installed on both sides of the interaction region. While low efficiencies are observed, of the order of a few percent, the resolutions of the two ToF detectors measured individually are 21 ps and 28 ps, yielding an expected resolution of the longitudinal position of the interaction, $z_{\mbox{\tiny vtx}}$, in the central ATLAS detector of $5.3 \pm 0.6$~mm. This is in agreement with the observed width of the distribution of the difference between $z_{\mbox{\tiny vtx}}$ measured independently by the central ATLAS tracker and by the ToF detector, of $6.0 \pm 2.0$~mm.
}
 
\author{The ATLAS Collaboration}
 
\AtlasRefCode{FDET-2019-06}
 
\AtlasJournal{JINST}
\AtlasJournalRef{JINST 19 (2024) P05054}
\AtlasDOI{10.1088/1748-0221/19/05/P05054}
\PreprintIdNumber{CERN-EP-2024-004}

\hypersetup{pdftitle={ATLAS document},pdfauthor={The ATLAS Collaboration}}
 
\begin{document}
\nolinenumbers

\maketitle
 
\tableofcontents

\section{Introduction}

A significant fraction of proton--proton ($pp$) collisions at the LHC proceeds via processes where one or both protons stay intact, i.e. $pp \rightarrow pX$ and $pp \rightarrow pXp$, respectively. When an intact proton is observed in the final state, it is typically accompanied by a large rapidity gap, a region devoid of hadronic activity~\cite{Dokshitzer:1987nc,Bjorken:1992er} between the final system $X$ and the leading proton(s). This topology is usually explained in terms of an absence of colour connection between the outgoing proton(s) and the final state $X$. Most of such processes are called diffractive and are described by Pomeron exchange~\cite{Pomeron,ParticleDataGroup:2022pth}.

Diffractive processes include elastic scattering ($pp \rightarrow pp$), central diffractive processes (CD, $pp \rightarrow pXp$), single diffractive dissociation (SD, $pp \rightarrow pX$), and double diffractive dissociation (DD, $pp \rightarrow XY$). The intact protons produced in SD and CD processes emerge deflected at very small angles relative to the incoming proton beam, such that they escape detection in the ATLAS inner detector and calorimeters. They can be detected and measured using dedicated proton detectors positioned far from the interaction point (IP) and close to the beam. These are known as forward proton detectors. In ATLAS~\cite{PERF-2007-01}, there are two forward proton detector systems, Absolute Luminosity for ATLAS (ALFA)~\cite{Khalek_2016} and ATLAS Forward Proton detector system (AFP)~\cite{ATLAS-TDR-24}. The scope of ALFA is to measure elastic processes and soft diffractive dissociation processes, requiring special beam optics settings, while AFP is used to collect data with forward protons during the nominal operation of the ATLAS detector to probe mainly hard-scale diffractive and photon-induced processes.
 
\begin{figure}[ttt!]
\centering
\includegraphics[width=0.8\textwidth]{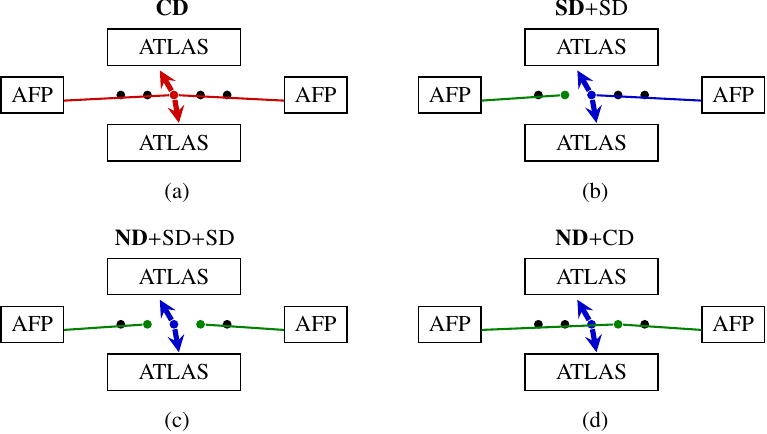}
\caption{Mechanisms of production of jet events with two forward protons: (a) central diffractive jets, (b) single diffractive jets + single diffraction, (c) non-diffractive jets + two single diffraction processes, (d) non-diffractive jets + central diffraction. Typical event vertices are indicated by dots. The primary processes detected by ATLAS are marked by vertices and arrows pointing to the central detector. The remaining vertices represent the pile-up processes.}
\label{fig:processes}
\end{figure}

At the LHC, measurements are greatly affected by the so-called pile-up effects.
Pile-up refers to independent proton--proton interactions that occur in the same bunch crossing and is quantified by the average number of interactions per bunch crossing, $\mu$.
For example, the presence of pile-up effects constitutes the primary obstacle to the observation of a large rapidity gap because particles produced in the additional interactions fill the gap.
 
In measuring diffractive processes with forward proton detection, the most important pile-up effect is the formation of a combinatorial background, produced by a coincidence of independent $pp$ interactions that together result in a signal-like event signature. For example, the signature of CD jet production is a pair of jets measured by the ATLAS calorimeter and two forward protons, one on each side of the IP, referred to here as a "double-tag".
However, the same signature can be obtained when combining two independent events, one with a non-diffractive jet pair and the other with a soft CD interaction, giving a forward proton on each side of the IP. Likewise, CD jet production can be mimicked by non-diffractive jet production recorded together with two soft SD processes, each giving an intact proton on one side, or by SD jet production together with a soft SD process. The scenarios described here are depicted in Figure~\ref{fig:processes}. These types of combinatorial background can be substantial because the cross sections for CD processes are much smaller than those for SD processes, which are in turn much smaller than for non-diffractive processes, when the same final state and the same hard scale are considered.

The AFP system consists of two detector stations on each side of the interaction point. In each station, the proton position is measured by a set of silicon tracking detectors, which are used to reconstruct the trajectory of the intact proton. Proton arrival times are measured in Time-of-Flight detectors (ToF) placed behind the farther of the two silicon tracking detectors. Each of these sets of detectors is contained within a Roman Pot, as first used at the CERN ISR~\cite{Fabjan:2312567}, whose horizontal movement from outside the LHC ring allows the detectors to be positioned as close as 2~mm from the nominal beam. More details are included in Section~\ref{sec:afpandtofdesign}.

The purpose of the ToF detector system lies in its ability to suppress the combinatorial background for processes with a double-tag by combining the time information from the two ToF detectors into the value of the expected production point (vertex) along the $z$-axis, $\zvtxkarel$. A comparison of this measurement with the precise $\zvtxkarel$ measurement from the ATLAS tracker provides a test of whether the protons were produced in the same place as the centrally detected charged particles. The level of combinatorial background suppression achieved by this test depends primarily on the timing resolution of the ToF detectors, see e.g.~\cite{Harland-Lang:2018hmi,Martins:2022dfg,ATLAS:2015dxo}. The role of ToF detector granularity, $\mu$-dependence and resolution is studied in Ref.~\cite{Cerny:2020rvp}, see also Ref.~\cite{Staszewski:2019yek}.
 
First ToF detector performance studies using data collected in 2017 can be found in Ref.~\cite{ATL-FWD-PUB-2021-002}. The final results presented in this paper include further improvements on the reported methods. In the first part, the measurement of efficiencies  (Section~\ref{sec:efficiency}) and timing resolutions (Section~\ref{sec:resolution}) of individual ToF channels is described, while the second part (Section~\ref{sec:analysisvtxmatching}) is devoted to tests of compatibility of the ToF measurements with the information from the central ATLAS detector: specifically, the $z$-coordinate of the beam spot and of the primary vertex. These studies provide a direct proof of concept of the ToF method for the selection of $pp\rightarrow pXp$ processes in ATLAS.

\section{Design of AFP and ToF detectors}
\label{sec:afpandtofdesign}
\color{black}
The AFP detector~\cite{ATLAS-TDR-24} consists of four stations, two stations on each side of the IP at 205 and 217~m, which are respectively denoted as NEAR and FAR. The sides of the ATLAS interaction region are denoted as A and C~\footnote{\AtlasCoordFootnote}, where side C corresponds to the side where the clockwise LHC beam leaves the interaction region, and vice versa for side A, which also coincides with the positive direction along the $z$-axis.
 
All four stations are equipped with silicon trackers (SiT) consisting of four layers of 3D silicon pixel detectors~\cite{Lange:2015xla}. The active area covered by each SiT is approximately $20 \times 20 \mbox{ mm}^{2}$ with a pixel size of $50 \times 250 \,\, \mu\mbox{m}^{2}$, forming a $336$ by $80$ pixel grid on each SiT plane. To improve the spatial resolution, all the SiT planes are tilted through an angle of $14^{\circ}$ around the vertical. This leads to resolutions of $6~\mu\mbox{m}$ and $30\,\mu\mbox{m}$ in the $x$- and $y$-coordinates, respectively, as measured in beam tests~\cite{ATLAS:2016lsc}.
 
\begin{figure}[ttt!]
\centering
\includegraphics[width=0.3\textwidth]{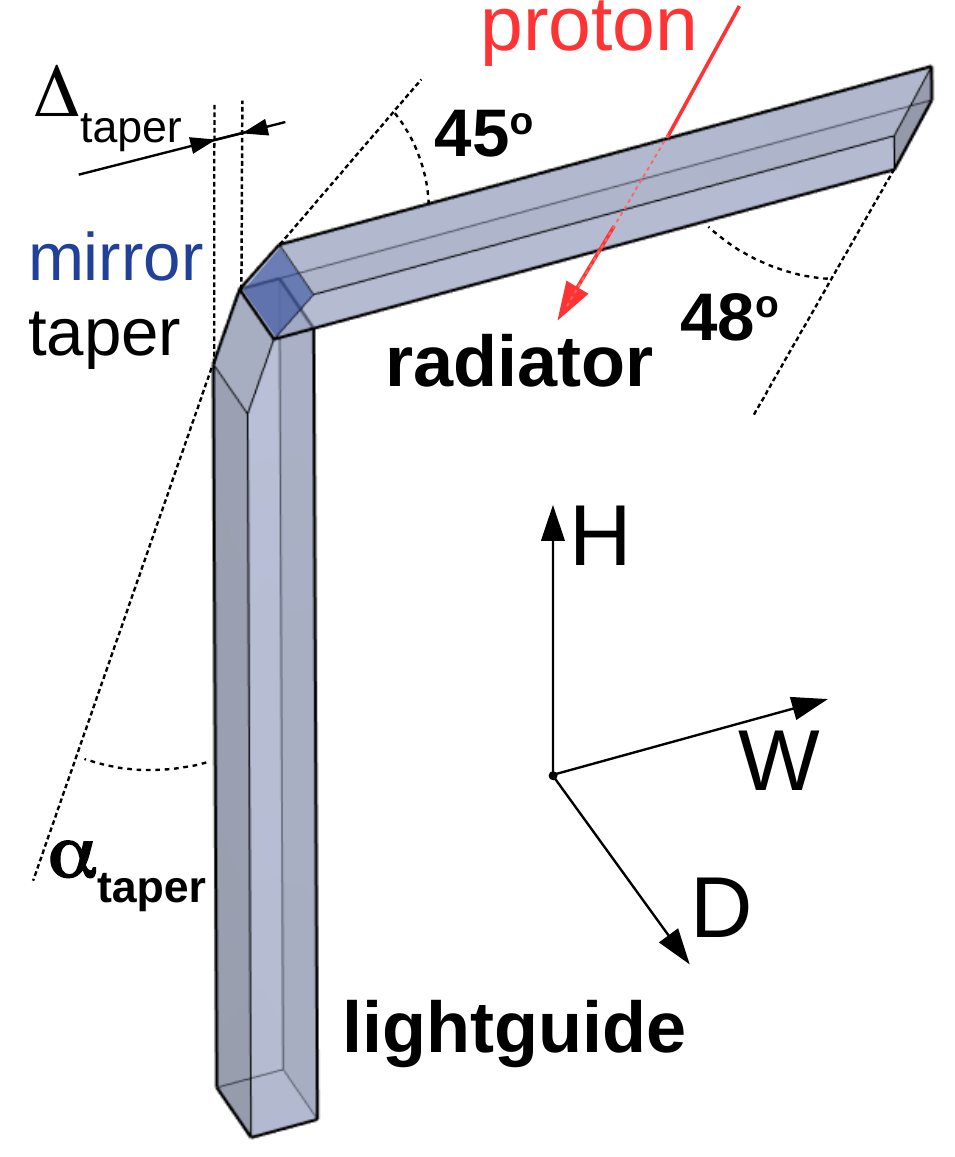}
\caption{ToF LQ-bar design. The shaded volumes represent quartz. The aluminium mirror is indicated by a blue surface in the LQ-bar elbow. The H, W and D (height, width and depth, respectively) directions refer to the local coordinate system of the LQ-bar used for specifications of dimensions in Table~\ref{tab:lqbardimensions}.}
\label{fig:lqbar}
\end{figure}
 
The ToF detectors are installed only in the FAR stations behind the trackers. The time-of-flight measurement is performed with fused silica~\cite{Suprasil} Cherenkov detectors forming a $4\times4$ matrix of L-shaped Quartz bars (LQ-bars)~\cite{Nozka:14}. The geometry of the ToF detector is such that the light yield is optimised within the space constraints of the Roman Pot stations. The LQ-bar consists of two arms: a radiator arm and a light-guide arm, see Figure~\ref{fig:lqbar}, where the two arms are glued together at 90 degrees by a UV-transparent epoxy glue. The geometric features of all LQ bars are detailed in Table~\ref{tab:lqbardimensions}. The photons emitted along the proton trajectory inside the radiator arm propagate into the light-guide arm, which is attached to a micro-channel-plate multi-anode photo-multiplier (MCP-PMT) with a 4x4 matrix of anode pads of similar performance as provided in~\cite{Photonis}. The radiator arms are oriented under the Cherenkov angle of $48^{\circ}$ with respect to the beam axis. This minimises the number of
photon reflections and the time needed for the light propagation through the bar. The elbow presents an aluminium mirror and a taper cut to achieve better focusing of the Cherenkov photons. The open end of the radiator arm  is cut parallel to the beam axis to reflect the photons back to the LQ-bar volume. This helps not only to increase the photon yield in each bar but also provides photon enrichment of subsequent bars in the direction of proton motion, as Cherenkov photons are produced with variable wavelengths, and their angles of total reflection can differ from the slant of the LQ-bar.

\begin{table}[ttt!]
\renewcommand{\arraystretch}{1.3} 
\caption{
Dimensions of the radiators and light guides of the ToF LQ bars. The sizes are specified in directions H, W and D as shown in the coordinate frame in Figure~\ref{fig:lqbar}. The absence of the taper in trains~2 and 3 is indicated by zero taper dimensions.
}
\small
\begin{center}
\begin{tabular}{|c|c|c|c|c|c|}
\hline
\multicolumn{6}{|c|}{LQ bar dimensions: $H \times W \times D$ [mm] / $\alpha_{taper}\,[{}^{\circ}]$ / $\Delta_{taper}$ [mm]} \\
\hline
Train & \multicolumn{1}{c|}{Radiators A} & \multicolumn{1}{c|}{Radiators B} & \multicolumn{1}{c|}{Radiators C} & \multicolumn{1}{c|}{Radiators D} & \multicolumn{1}{c|}{Light-guides}\\
 
\hline
0 & $2 \times 62.41 \times 6 $ & $2 \times 56.78 \times 6 $ & $2 \times 51.15 \times 6 $ & $2 \times 45.52 \times 6 $ & $71.3 \times 5 \times 6 / 18 / 3$ \\
1 & $4 \times 58.16 \times 6 $ & $4 \times 52.53 \times 6 $ & $4 \times 46.9 \times 6 $ & $4 \times 41.27 \times 6 $ & $67.2 \times 5 \times 6 / 18 / 1$ \\
2 & $5 \times 52.91 \times 6 $ & $5 \times 47.28 \times 6 $ & $5 \times 41.65 \times 6 $ & $5 \times 36.02 \times 6 $ & $62.1 \times 5 \times 6 / 0 / 0$ \\
3 & $5.5 \times 46.6 \times 6 $ & $5.5 \times 43.03 \times 6 $ & $5.5 \times 35.4 \times 6 $ & $5.5 \times 29.77 \times 6 $ & $56.6 \times 5.5 \times 6 / 0 / 0$ \\
\hline
\end{tabular}
\end{center}
\label{tab:lqbardimensions}
\end{table}
 
\begin{figure}[tbh!]
\centering
\includegraphics[width=0.65\textwidth]{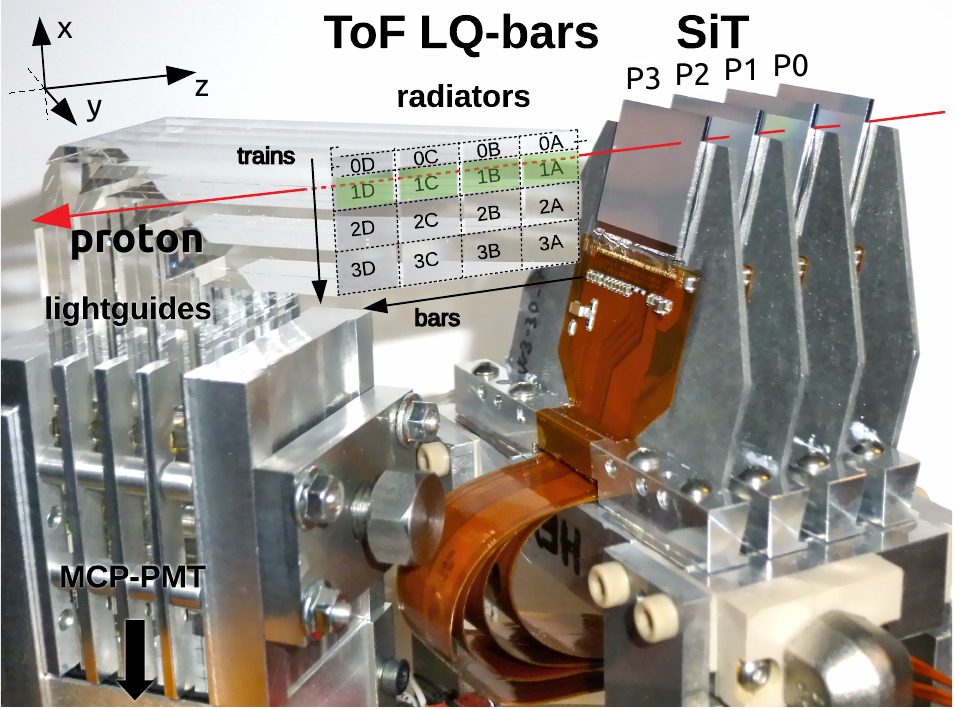}
\caption{Assembled AFP detector composed of the Silicon tracker and the Time-of-Flight detector. The leading proton trajectory is indicated with an oriented red line. The Roman Pot flange is located below the region depicted in the figure. The indicated coordinate system coincides with the one in Figure~\ref{fig:flangeCAD}. The segmentation of the ToF into trains and channels is indicated and the channels of the train traversed by the proton are highlighted.}
\label{fig:sittof}
\end{figure}

\begin{figure}[tbh!]
\centering
\includegraphics[width=0.75\textwidth]{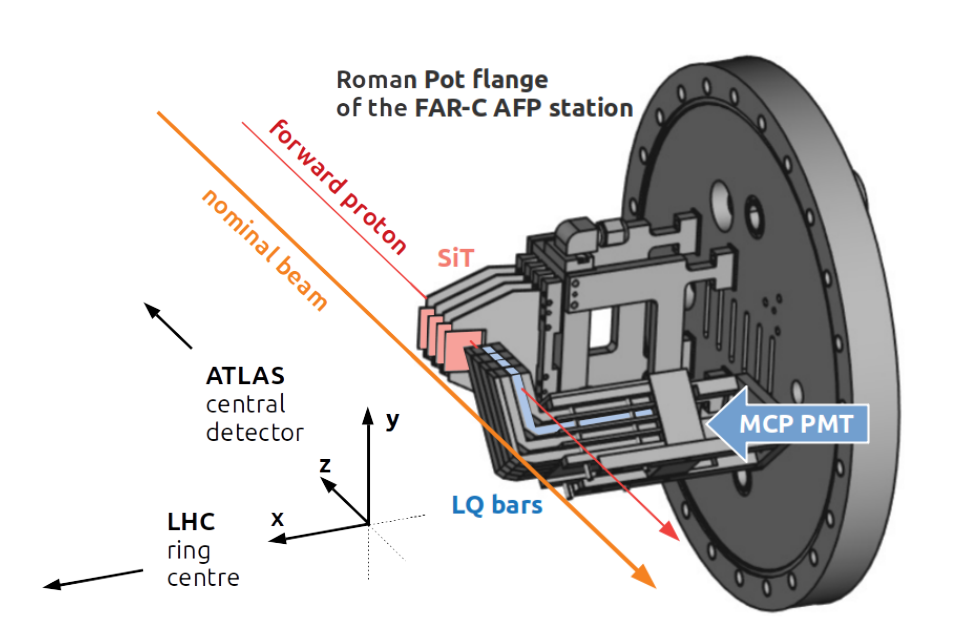}
\caption{Schematic view of the AFP FAR-C station. For clarity, only the Roman pot flange without the cover is shown. The set of LQ-bars attached to the MCP-PMT is fixed in a holder which (together with the holder for the SiT) is mounted to the flange. The LQ bars traversed by the leading proton (shown as a red arrow) are highlighted in blue colour. Also the proton passage through the SiT tracker is indicated. The position of the nominal beam is shown by an orange arrow. The local right-handed reference frame is defined such that the positive $x$-axis is perpendicular to the flange and points to the LHC ring centre, the $y$-axis is perpendicular to the LHC ring plane and points upwards, and the positive $z$-axis represents the anticlockwise direction tangential to the LHC circumference pointing to the ATLAS IP.}
\label{fig:flangeCAD}
\end{figure}

Four bars, labelled A -- D,  are placed sequentially in the beam direction to form a train, bar~A being the first to be crossed by protons. There are four trains in each ToF detector, numbered from 0 to 3 as the distance from the beam increases. The optical path in all bars is equalised by employing radiator arms of decreasing length along the direction of motion of the protons. Figure~\ref{fig:sittof} shows a photograph of the assembled ToF and SiT mounted on the Roman pot flange while Figure~\ref{fig:flangeCAD} presents the schematic view of the leading proton detection in the FAR-C AFP station.
 
The number of Cherenkov photons emitted in a bar determines the number of photo-electrons released in the corresponding PMT photo-cathode via its quantum efficiency and was estimated to be in the range between 20 and 40 photo-electrons per channel. The high voltage (HV) applied enables the photoelectrons that enter the pores of the MCP to initiate an electron avalanche that is collected at the nearest read-out anode. The voltage pulse registered at the anode undergoes a further two-stage amplification and is processed by a constant-fraction discriminator (CFD~\cite{CFD}). The CFD produces a square signal, whose start (also referred to as a timestamp) and duration are determined by the moments when the pulse is above a predefined fraction of its pulse height. The CFD output is sent to a high-performance time-to-digital converter (HPTDC~\cite{HPTDC}). The HPTDC samples the $25$~ns time intervals between consecutive bunch crossings into $1024$ bins. All the stages of signal formation and processing contribute to the final timing resolution.

\section{Performance of single channels}
\subsection{Data and event selection}
Analyses of the AFP data that do not require information from the central detector can profit from the data recorded in the so-called AFP calibration stream. These data sets are suitable for performance studies because they contain only events with AFP information.
The analysis of the performance of single ToF channels is based on the following five sets of ATLAS data recorded in 2017: $331020$, $336505$, $341419$, $341534$ and $341615$ using the numbering of ATLAS runs.
As documented in Table~\ref{tab:runs}, these runs provide large data samples with low levels of pile-up. The majority of data in 2017 were taken at higher $\mu$ and the AFP-event triggers were substantially prescaled. As a consequence, the statistics available for each of these runs were limited, and are thereby not expected to improve the single-channel performance results significantly. No effort is made to consolidate the bulk of the high $\mu$ 2017 data into a unified dataset.
 
The primary event selection relies on local AFP triggers, which can be based either on the SiT or ToF information. For the data analysed here, the SiT-based trigger is used.
The triggering decision is based on the presence of a signal in at least two out of the first three layers of the SiT at the given AFP station. This makes possible a measurement of the ToF efficiencies by using samples of events selected by requiring local tracks in the SiT in the FAR stations in front of the active volumes of the LQ-bars.
 
The arrival times of protons emerging from $pp$ collisions are anticipated to fall within well-constrained time intervals of 2.5~ns at most, significantly narrower than the period between the consecutive bunch crossings of 25~ns. However, in certain ToF channels the measured distributions of the arrival times were sometimes significantly broader and higher in the entire range of 25~ns. Such lumiblocks (LB) are removed from further analysis. Lumiblocks are conventional time intervals for recording ATLAS data, and last typically one minute. The rejected data represent only a negligible fraction of the dataset, well below one percent in the affected runs. The most likely reason for this effect is a temporary increase of non-collision background levels.

Examination of the detector hardware conducted after the 2017 data taking revealed that malfunctioning first-stage preamplifiers of channel A in train 0 on side A and channel D in train 0 on side C caused these two channels to appear as inactive during 2017.
 
\begin{table}[htb!]
\caption{List of AFP calibration stream runs recorded in 2017 and used for the efficiency and time resolution studies. The HV column indicates the high voltage applied to the MCP-PMTs.}
\begin{center}
\begin{tabular}{|c|c|c|c|c|c|c|}
\multicolumn{1}{c}{Run} & \multicolumn{1}{c}{Date}  & \multicolumn{1}{c}{AFP trigger} & \multicolumn{1}{c}{ToF HV~{[}V{]}} & \multicolumn{1}{c}{$N_{\mathrm evt}$} & \multicolumn{1}{c}{$\mu$} & \multicolumn{1}{c}{${L}_{\mathrm int.}$~[pb${}^{-1}$]}\\
\hline
331020  & July 29       &  SiT              & -2000                  &  45M           &  $\sim 1$ & 14 \\
336505  & Sept 23       &  SiT              & -2000                  & 143M           &  $\sim 0.04$ & 17\\
341419  & Nov 22       &  SiT              & -1950                  & 240M           &    $\sim 2$ & 31\\
341534  & Nov 23       &  SiT              & -1950                  & 430M           &    $\sim 2$ & 51\\
341615   & Nov 25       &  SiT              & -1950                  & 240M           &    $\sim 2$ & 31\\
\hline
\end{tabular}
\end{center}
\label{tab:runs}
\end{table}

\subsection{ToF noise}
The hits registered in the ToF channels not initiated
by the passage of the leading proton emerging from the $pp$ collision will be referred to as ToF noise. Two types of sources of ToF noise (random and beam-related) are discussed, and will be elaborated upon in the subsequent sections.

\subsubsection{Random ToF noise}
There are at least two sources of random noise in the ToF detectors. The first is caused by a thermal liberation of electrons inside the PMT (either from a photo-cathode or from MCP layers) and is called dark pulses. A conservative estimate based on PMT data-sheet values gives a small $\sim 0.1\%$ dark pulse occurrence probability inside the 2.5~ns range of expected proton arrival times per channel. Another potential source of time-uniform noise may arise from PMT pulse fluctuations (e.g. from interference with external electromagnetic fields), accepted by the CFD. In general, any random signals measured by the ToF detector are easily recognisable since they are expected to be distributed uniformly over the $25$~ns measurement range. We do not observe such an additional uniform distribution, from which we conclude that the probability of accidental ToF hits occurring without any incident particle is negligible.

\subsubsection{{Non-collision background}}
\label{sec:BIB}
 
The ToF detectors are sensitive to charged particles that are rapid enough to produce Cherenkov light in the bars. There is a  component that is neither noise nor caused by the detection of leading protons from $pp$ interactions; this is commonly referred to as non-collision background. It is mainly due to the passage into the ToF detector of particles originating in interactions of beam protons with beam-pipe collimators or residual gas, leading to the production of secondary particles which may further interact with the beam pipe material -- all this is commonly referred to as the beam halo. Also, secondaries from showers generated by the interaction of leading protons with the Roman Pot walls or floor in the NEAR stations are expected to be present in the data. Since they arise directly or indirectly from beam protons, these  secondaries are synchronous with the LHC beam bunches.
 
Depending on the LHC filling scheme, not all the bunches are filled. The bunch crossings (denoted by bunch crossing ID, BCID) are labelled as {\bfseries paired} in the case when two populated bunches from the two beams cross, {\bfseries empty} if none of the physics bunches is filled and {\bfseries unpaired} if only one of the bunches is filled. These BCID scenarios give access to data samples between which the different contents of non-collision background can be compared.
 
\begin{figure}[tb!]
\centering
\subfloat[]{\includegraphics[width=0.44\textwidth]{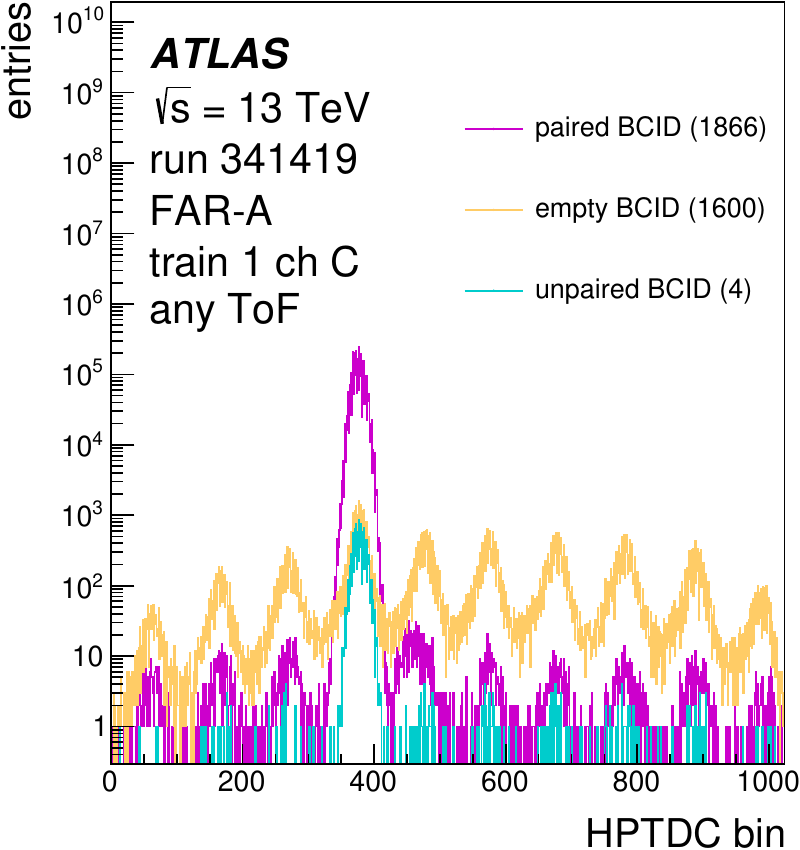}}\hspace{0.2cm}
\subfloat[]{\includegraphics[width=0.435\textwidth]{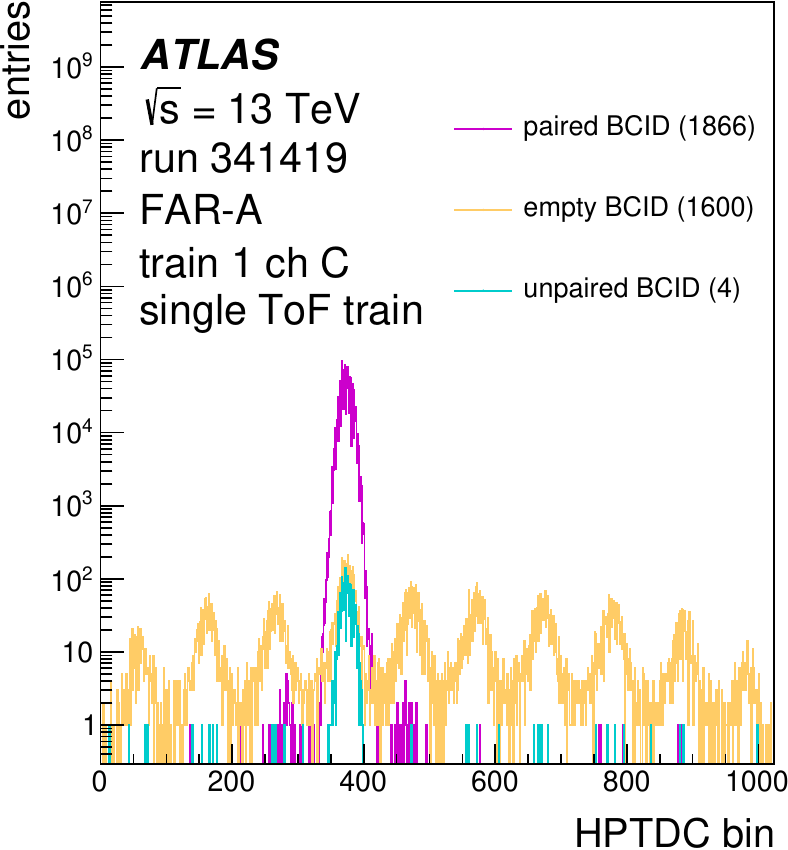}}
\caption{Raw-time distributions measured in paired, empty and unpaired BCID events in the FAR-A station in channel~C of train~1 for the any-ToF selection (left panel) and the single-ToF-train selection (right panel). The displayed range corresponds to the entire time period of 25~ns. The numbers in brackets indicate the number of contributing BCID cases to each particular BCID scenario.}
\label{fig:rawtimesBIB341419}
\end{figure}

Figure~\ref{fig:rawtimesBIB341419} shows the raw-time distributions collected in channel C and train~1 of the FAR-A station measured in run 341419. The distributions are obtained without any selections applied at the level of the SiT apart from those imposed by the trigger. The left panel shows the raw-time distribution in the particular channel irrespective of whether any other ToF channels contained hits ({\bfseries any-ToF}). The distributions are shown for the paired, empty and unpaired BCID cases. The range of arrival times of diffractive protons originating in the paired BCID cases is clearly visible as an enhancement centered around the HPTDC bin number 380. The exact position of this range is an artefact of time delay settings of the ToF readout with respect to the LHC clock.
The values outside the expected time range correspond to halo activity initiated by satellite bunches. These are present due to the fact that the radiofrequency (RF) accelerator cavities operate at 400~MHz which divides the LHC beam longitudinally into so-called RF buckets of 2.5~ns length. Only one of the ten RF buckets is reserved to contain the beam protons for collisions.
The other RF buckets, may contain protons originating from spill-over processes (so called satellite bunches). The raw-time distributions recorded in the cases of empty BCID (left panel of Figure~\ref{fig:rawtimesBIB341419}) show that the ToF detector is sensitive to the halo activity caused by the satellite bunches. The structure of ten 2.5~ns RF buckets is well pronounced. Eventually, the unpaired BCID raw-time distributions indicate that the mere passage of bunches rich in protons
can lead to time measurements consistent with the expected ones. This means that a certain fraction of the ToF hits can be caused by non-collision events in the expected time range even in the paired BCID events.
 
The distributions in the right panel of Figure~\ref{fig:rawtimesBIB341419} are obtained under the condition that the ToF hits are observed exclusively in a single train, which will be referred to as a {\bfseries single-ToF-train} selection throughout this document. The distributions illustrate that the single-ToF-train requirement leads to a suppression of measured times outside the expected range. This is caused by the removal of events where ToF hits are spread across the trains and channels, thus, preferring the (by design) expected ToF hit topologies created by diffractive protons traversing LQ-bar radiators of one train only.
Note that the selection of single-ToF-train is effective not only in suppressing non-collision background but also in rejecting high-multiplicity ToF hit events, particularly in cases when genuine diffractive protons generate hadronic showers.
At the same time the single-ToF-train requirement leads to a reduction of the event yield in the expected time range by about a factor of two for the paired BCIDs and a factor of ten otherwise, as can be seen from Figure~\ref{fig:rawtimesBIB341419}. This can be explained by the fact that the empty BCID events contain the halo activity more frequently and thus are more efficiently suppressed by the requirement of single-ToF-train signal topology.

\color{black}
Another handle to assess the non-collision background is provided by the SiT. A data-driven technique was developed based on the properties of spatial distributions of tracks reconstructed in the SiT in terms of local $x$ and $y$ positions. The data from the paired BCIDs (dominated by diffractive protons) show a clear $x$--$y$ correlation, while those from the empty BCIDs (dominated by background) are scattered over the whole $(x,y)$-plane. This difference enables us to estimate the magnitude of the background contribution in the analysed data, which is observed to decrease with decreasing SiT track multiplicity. The effect of the track multiplicity cuts on the background suppression is presented in Section~\ref{sec:analysisvtxmatching}.
 
\FloatBarrier
 
\subsection{Measurement of the efficiency}
\label{sec:efficiency}
The efficiencies of the ToF detectors are measured at the level of trains using a sample of events where exactly one SiT track is reconstructed. The efficiency of a single ToF channel is calculated as
\begin{equation}
\varepsilon_{ijk} = \frac{N(\mbox{bar-ij} \,\, \cap \,\, \mbox{track-k)}}{N(\mbox{track-k})},\nonumber
\end{equation}
where $N(\mbox{bar-ij} \,\, \cap \,\, \mbox{track-k})$ represents a number of events with a signal in the ToF bar-channel $i$ of the train number $j$ in the sample of events with SiT tracks in geometrical acceptance of the train number $k$, see Figure~\ref{fig:sittof} for clarity. In this way, $N($track-k$)$ is the total number of events with SiT tracks in the geometrical acceptance of the train number $k$. Note that the distinction between the indices $j$ and $k$ means that the efficiency can be evaluated also in cases where the SiT tracks do not point to the geometrical acceptance of the train whose channels are being evaluated.
 
\begin{figure}[t!]
\centering
\subfloat[]{\includegraphics[width=0.49\textwidth]{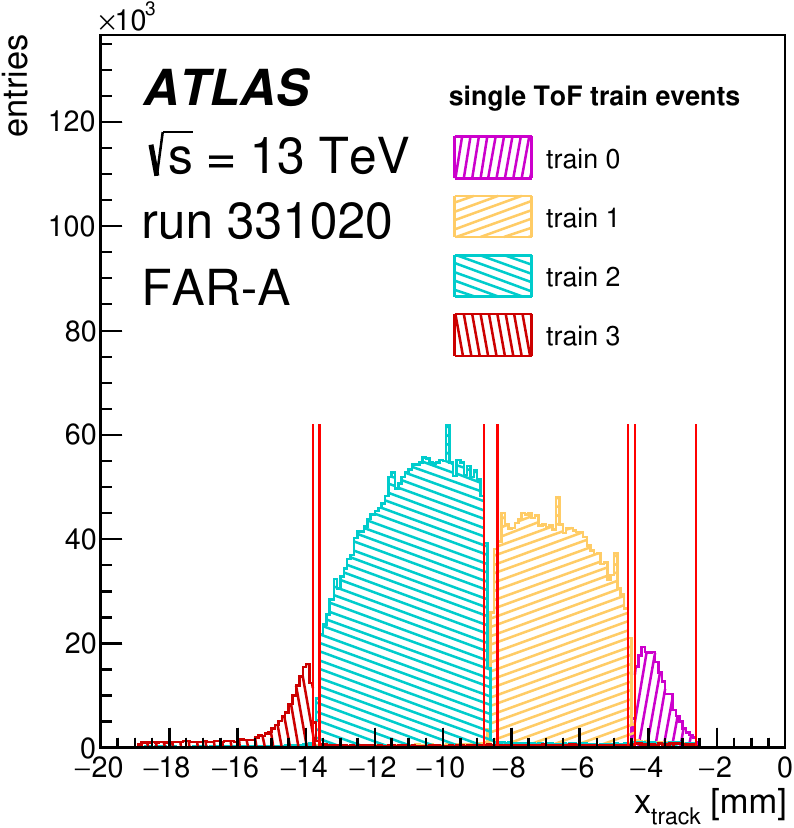}}
\subfloat[]{\includegraphics[width=0.49\textwidth]{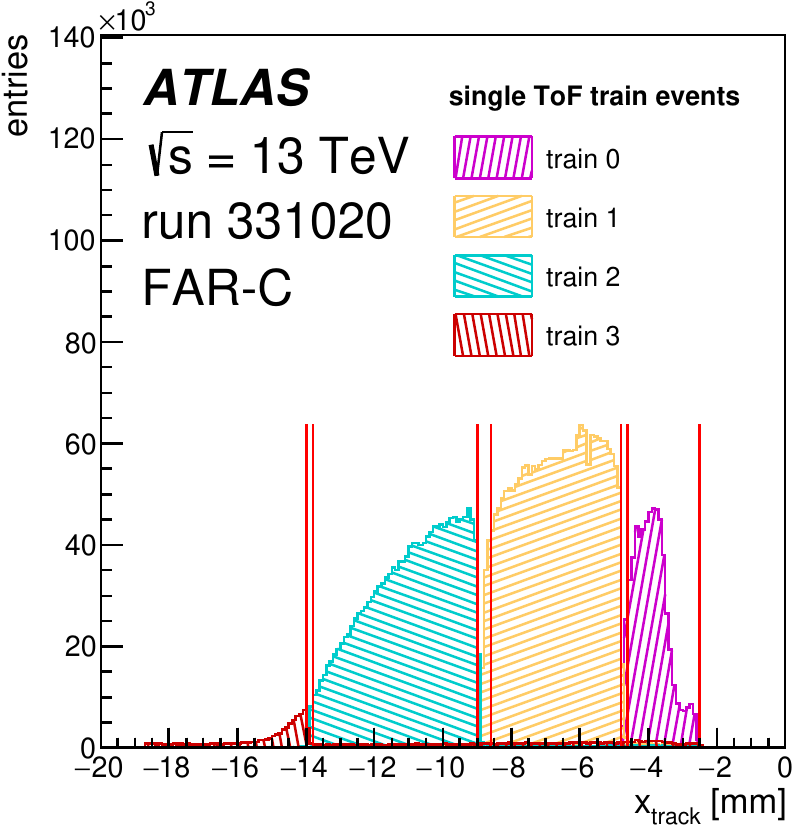}}
\caption{Distributions of local track $x$-coordinates in the AFP-FAR station after applying the single-ToF-train selection in the AFP calibration stream run $331020$. The red vertical lines indicate the chosen cuts that are used to define the acceptance of the trains. The left and right plots correspond to the A (a) and C (b) side, respectively.}
\label{fig:ROIs331020}
\end{figure}

The geometrical overlap of the SiT tracks with the ToF trains is determined by looking at local track $x$-coordinate distributions saved under the condition that channels of only a single train are hit, see Figure~\ref{fig:ROIs331020}. In this way, the ranges of $x$-positions of the tracks that correspond to the geometrical acceptance of the individual trains are identified.

The detection efficiency of the ToF train can be evaluated as
\begin{equation}
\varepsilon_{jk} = \frac{N\left[\left(\mbox{bar-Aj}\cup\mbox{bar-Bj}\cup\mbox{bar-Cj}\cup\mbox{bar-Dj}\right) \,\, \cap \,\, \mbox{track-k}\right]}{N(\mbox{track-k})},
\label{eq:traineff}
\end{equation}
where for each train $j$ the condition is a simple requirement of a hit observation in any channel of this train.
 
\begin{figure}[tbh!]
\centering
\subfloat[]{\includegraphics[width=0.48\textwidth]{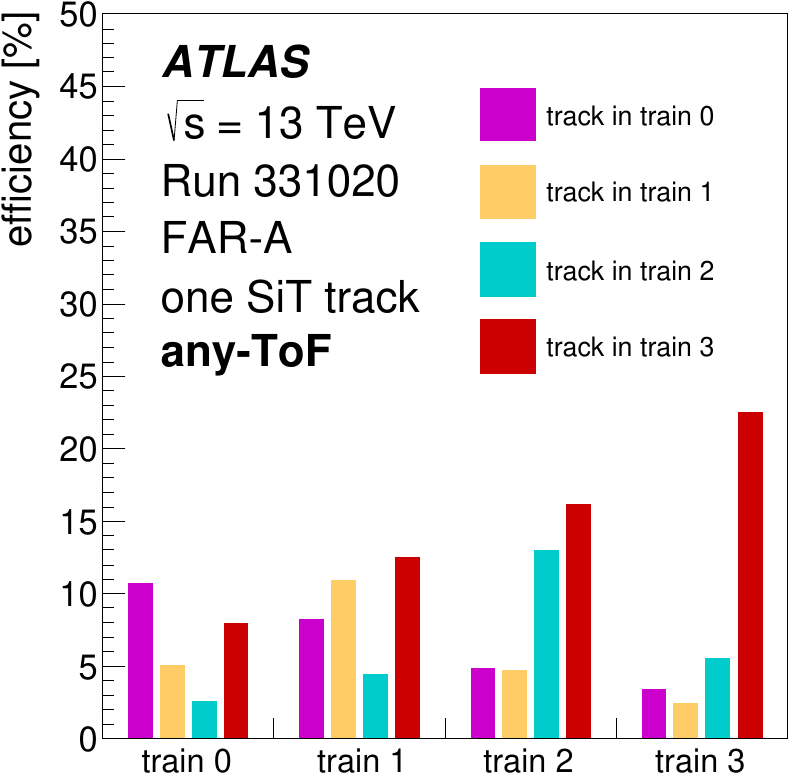}}\hspace{0.3cm}
\subfloat[]{\includegraphics[width=0.48\textwidth]{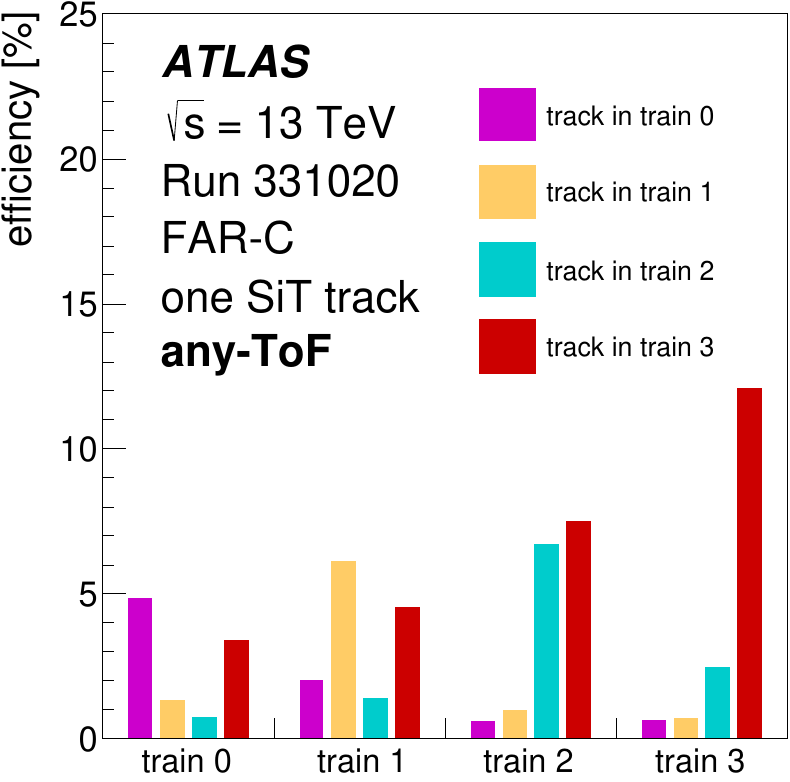}}
\caption{Efficiencies of ToF trains obtained from the AFP calibration stream data corresponding to ATLAS run 331020 in the A-FAR (a) and C-FAR (b) AFP stations. The data are required to contain exactly one reconstructed SiT track with no further constraints applied in the ToF (the so called any-ToF selection). The relative statistical uncertainties are always by two or more orders of magnitude smaller than the actual efficiency values and are not displayed.
}
\label{fig:331020traineffany}
\end{figure}

In Figure~\ref{fig:331020traineffany} the train efficiencies are shown in run 331020 measured, using Equation~\ref{eq:traineff}, under the condition of a local SiT track reconstructed in only one of the four train acceptance regions. No further requirement is applied on the position of the ToF hits (any-ToF). For the events with the track pointing to a given train (corresponding to the bars with the same colour in Figure~\ref{fig:331020traineffany}), the maximum efficiency is always obtained for the train to which the SiT track is pointing. However, the maximum is not very peaked and large efficiencies are also observed for the neighbouring trains. This can be explained by the presence of secondary particles created in front of the ToF detector traversing the ToF LQ-bars. The train efficiencies measured for the trains to which the tracks are pointing (track-matched trains) vary between 10--20\% and 4--12\% in the FAR-A and C stations, respectively. The train efficiencies in the non-track-matched trains are always lower and decrease with the distance from the track-matched train. This decrease is weaker in the case of tracks reconstructed in train 3 of both stations, indicating possibly a large influence of secondaries from hadronic showers generated at collimators located in front of the Roman Pots.
 
\begin{figure}[tbh!]
\centering
\subfloat[]{\includegraphics[width=0.48\textwidth]{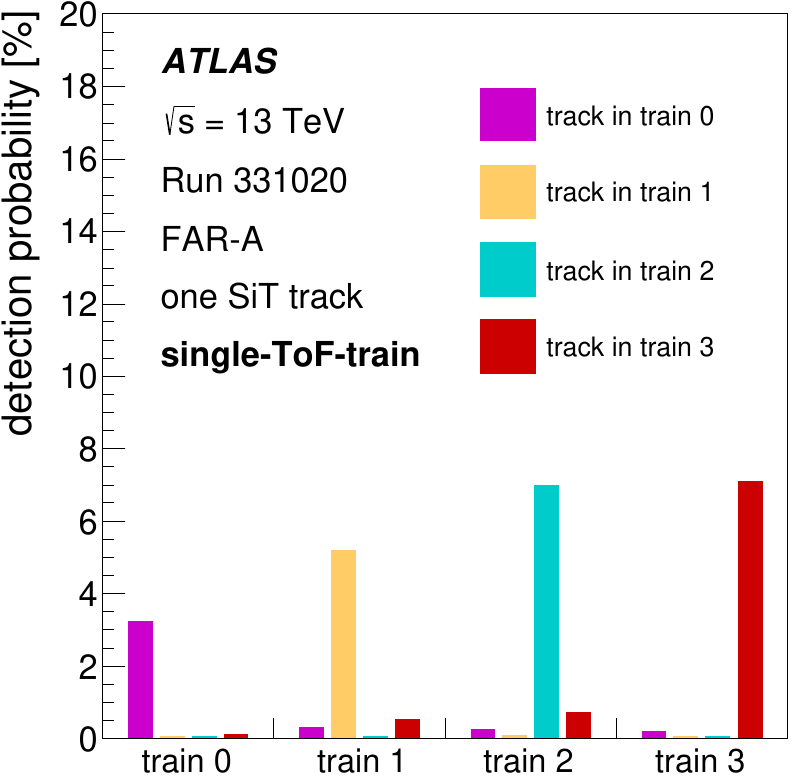}}\hspace{0.3cm}
\subfloat[]{\includegraphics[width=0.48\textwidth]{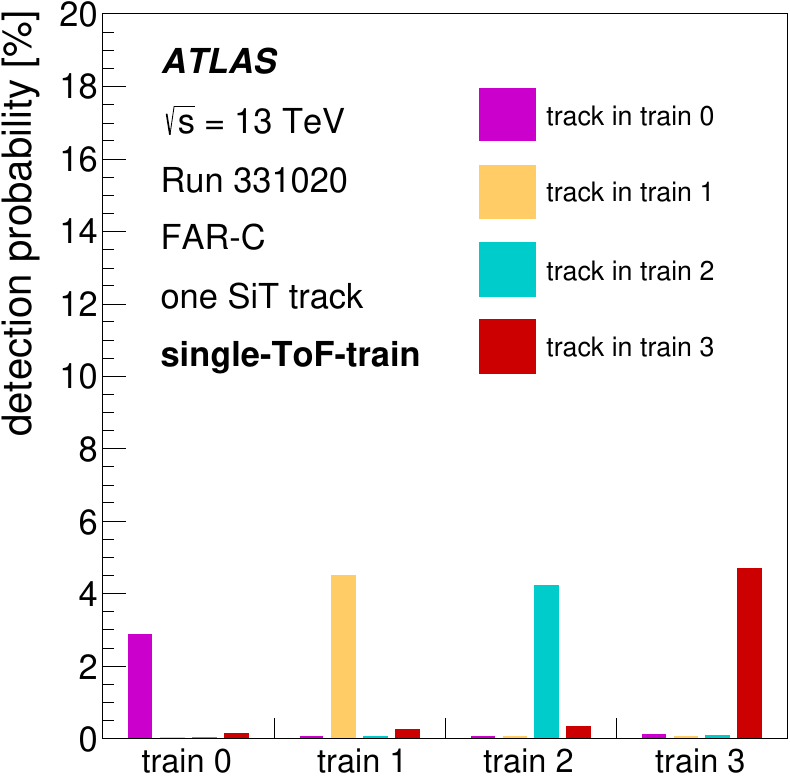}}
\caption{The detection probabilities of the ToF trains measured in the sample of events with one reconstructed SiT track utilising an additional constraint of ToF channel hits being in the single-ToF-train topologies, presented in the same manner as in Figure~\ref{fig:331020traineffany}.}
\label{fig:331020traineffclean}
\end{figure}

Figure \ref{fig:331020traineffclean} presents the probabilities of observing a single-train ToF signature provided a single SiT track is reconstructed. A decrease in probabilities of signal detection in the ToF trains by a factor between two and three on both sides is observed.
The most efficient trains are those pointed by the track, while only a small fraction of events is registered in the neighboring trains.
This observation indicates that the single-ToF-train condition represents a selection cut that mitigates the effects of secondaries from particle showers.

Quantitatively similar efficiencies to run 331020 are measured in run 336505, whereas for the late 2017 runs 341419, 341534 and 341615 a substantial drop of efficiencies in every ToF train is observed. The typical values measured on the FAR-A side are around 2--4\%, while the FAR-C side ones decrease by another factor of ten to a 0.2--0.4\% level, where the efficiency decrease is partially caused also by the fact that the MCP-PMTs were operated at lower HV. The asymmetry between the efficiencies measured on the A and C sides persists for all analysed data. It can be related to different levels of the beam-halo background present on the two sides.

The overall low detection efficiencies are attributed to MCP-PMTs exceeding expected lifetimes, which were known to be of the order of 0.5~C/cm$^{2}$ of integrated charge. The calculations based on using conditions from the 2017 data taking indicate that the actual exposure of the PMTs was at least 10~C/cm$^{2}$.
The MCP-PMT types that are designed for longer lifetimes, making use of the Atomic Layer Deposition (ALD) technique, were not available for the Run 2 data taking. Their integrated charge tolerance is two orders of magnitude higher~\cite{Lehman1} and, among other improvements documented in~\cite{Nozka:23}, they were chosen for the operation of the ToF detector in Run 3.

\subsection{In situ timing calibration of HPTDC}
\label{hptdc}
The HPTDC performs the time measurement by dividing the time window of $25$~ns between the consecutive bunch crossings at the LHC into 1024 bins. In the case of an ideally calibrated HPTDC, the width of raw-time bins would be equal to 25/1024~ns~($\sim 24.4$~ps). While in principle possible, the intrinsic calibration of HPTDCs was not performed for the studied data period. The actual raw-time bin widths of this uncalibrated device are therefore non-trivial and they need to be detemined. The circuitry behind the HPTDC introduces repeating patterns (nearly regular modulations) in the raw-time distribution which can be suppressed later, during the offline analysis of the data, using the Fast Fourier Transform (FFT).
 
\begin{figure}[tb]
\center
\subfloat[]{\includegraphics[width=0.48\textwidth]{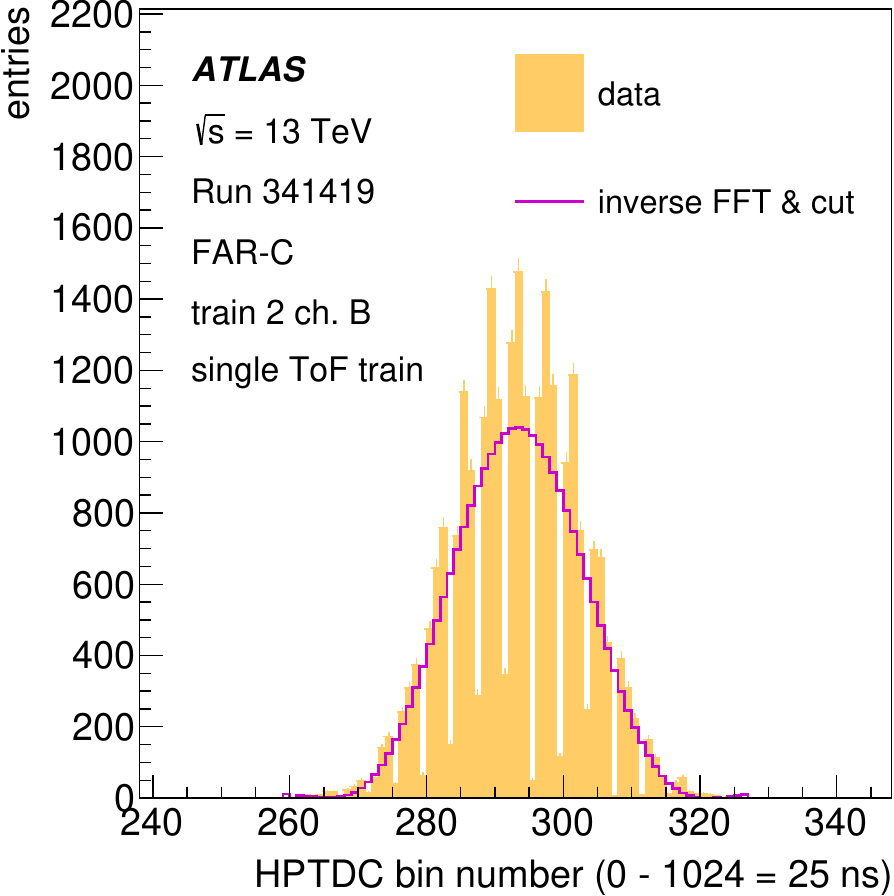}}\hspace{0.3cm}
\subfloat[]{\includegraphics[width=0.49\textwidth]{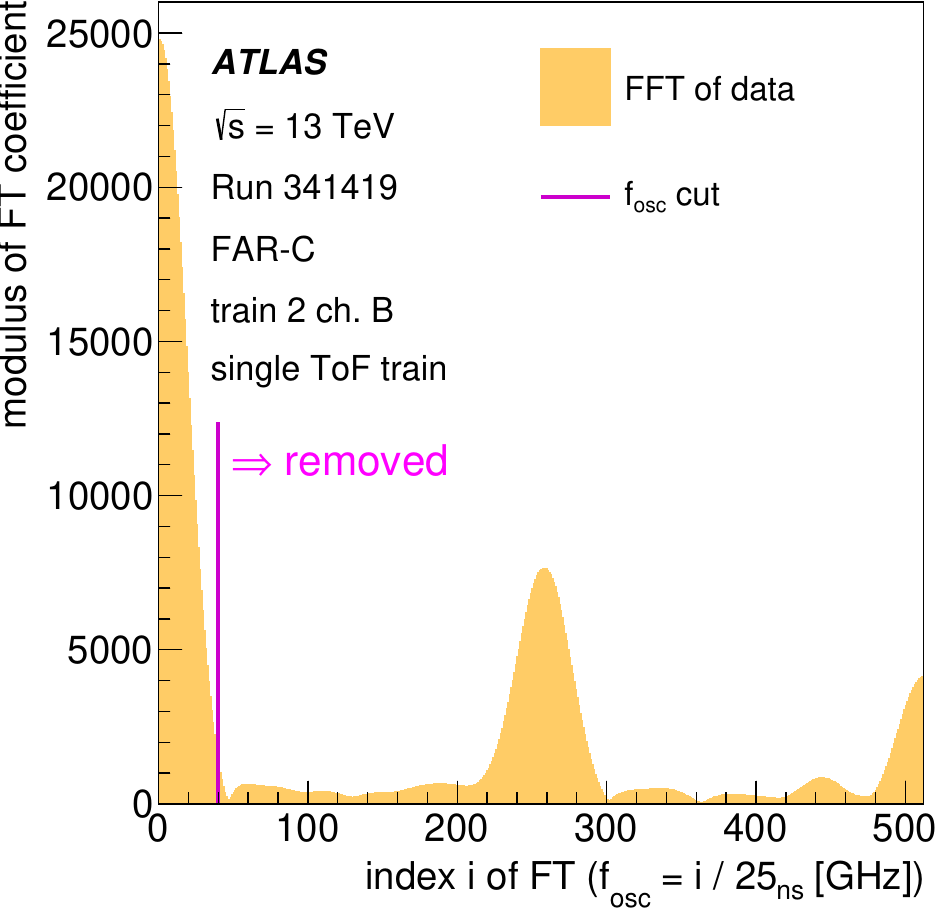}}
\caption{HPTDC calibration plots for channel~B of train~2 in  station FAR-C obtained from the AFP calibration run 341419. (a) The raw-time distribution (filled histogram), the inverse Fourier transform (line histogram) with suppressed oscillations above 1.6~GHz. (b) Distribution of the moduli of the FFT coefficients. The distribution is symmetric around the maximum displayed index value of 512 extending up to index 1023. The vertical line in the plot denotes the suppressed oscillation frequencies for indices between 40 and 984, a consequence of the plot symmetry. This symmetry arises from the chosen convention for the complex phase range, which spans from 0 to 2$\pi$.}
\label{fig:hptdccalib341419-1}
\end{figure}

\begin{figure}[hhh]
\centering
\includegraphics[width=0.49\textwidth]{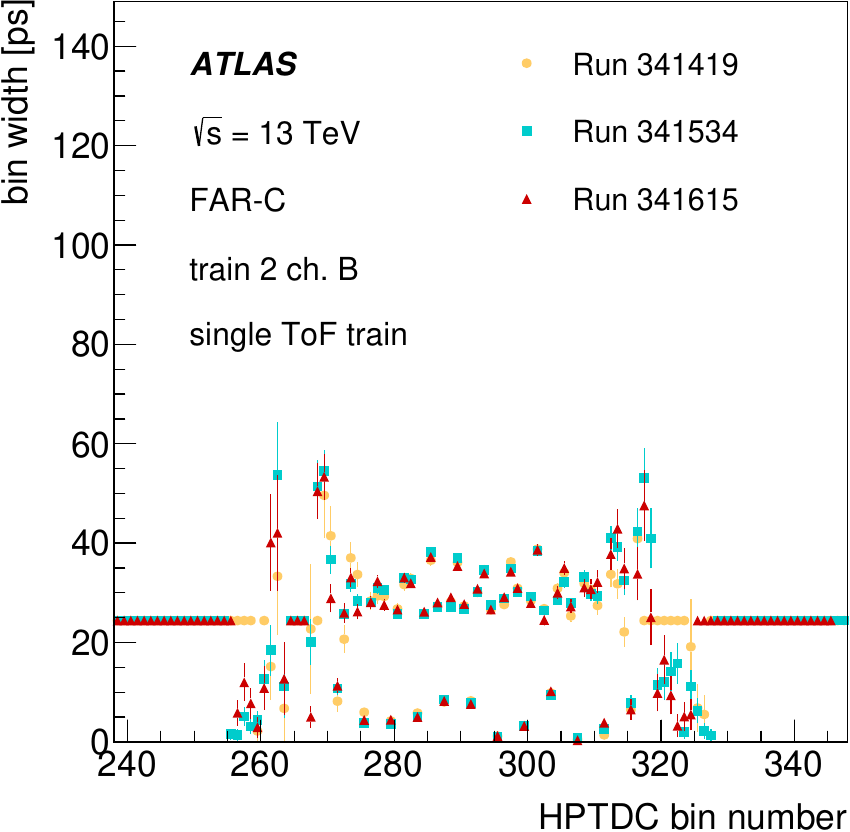}
\caption{Comparison of bin widths extracted in channel~B of train~2 in the FAR-C station for runs 341419, 341534 and 341615.}
\label{fig:hptdccalib341419-2}
\end{figure}

The oscillations of the raw-time distributions are removed by using a simple cut on the maximum allowed frequency (or minimum length of oscillations) in the distribution of moduli of the FFT coefficients of the original raw-time distributions, as illustrated in Figure~\ref{fig:hptdccalib341419-1} for channel B of train 2 in station FAR-C. Oscillation periods shorter than 26 raw bins (corresponding to about 635~ps or roughly to 1.6~GHz) are filtered away, and the ratio of the original raw-time distribution to the inverse FFT is used to evaluate the actual HPTDC bin width. The bin widths are compared between runs 341419, 341534 and 341615, see Figure~\ref{fig:hptdccalib341419-2}, and are found to be consistent at the level of single ps in regions with sufficient statistics. The bin widths from low statistics bins are determined with sizeable statistical uncertainty and are assigned the nominal bin width. Once the bin widths are known, the calibration of the raw-times is done by calculating the shift of the measured bin position due to the non-trivial widths of the preceding bins. The run-to-run differences between the extracted widths observed in the tails of the raw-time distributions, lead unavoidably to artificial shifts of the bin positions. Such shifts are, however, of no concern since they are later absorbed into a constant time delay of each channel in every run individually. This means that the HPTDC bin centre corrections are unique for each run.
 
With the knowledge of the bin widths, a uniform smearing around the given bin centre can be applied.
This procedure provides the possibility to smooth the measured channel times as well as their differences, which helps to mitigate the artefacts related to the HPTDC binning. The time differences are quantities central to analysis of single-channel timing resolutions and also to the use of the ToF method for vertex reconstruction.
\color{black}
 

\FloatBarrier

\subsection{Measurement of single-channel time resolutions}
\label{sec:resolution}
 
This section describes the procedure used to extract the time resolutions of individual ToF channels.
The time measured in an $i$-th ToF channel consists of the following contributions:
\begin{equation}
t_{i} = t_{\mbox{\tiny proton}} + t_{i,\mbox{\tiny delay}} + t_{i,\mbox{\tiny smear}} - t_{\mbox{\tiny clock}},
\label{eq:channeltime}
\end{equation}
 
where $t_{\mbox{\tiny clock}}$ represents the signal of the LHC reference clock that opens the window of $25$~ns in the HPTDC inside which the leading protons from a single bunch crossing arrive at the ToF detector and is thus common to all channels on both sides. The $t_{\mbox{\tiny proton}}$ represents the proton arrival time, whose event-by-event variations with respect to $t_{\mbox{\tiny clock}}$ are determined by the properties of the luminous region (the so-called beam-spot), spanning usually over few hundreds of ps. The $t_{i,\mbox{\tiny delay}}$ is a constant channel time offset caused by a signal delay (for instance signal cable lengths, a global offset of HPTDC with respect to the LHC clock and the HPTDC bin centre corrections discussed in the previous section). The $t_{i,\mbox{\tiny smear}}$ represents all random aspects of the signal processing such as a variation in Cherenkov photon (photo-electron) statistics and effects of electronics, and whose width is used to address the timing resolution.
 
\begin{figure}[tb!]
\centering
\subfloat[]{\includegraphics[width=0.46\textwidth]{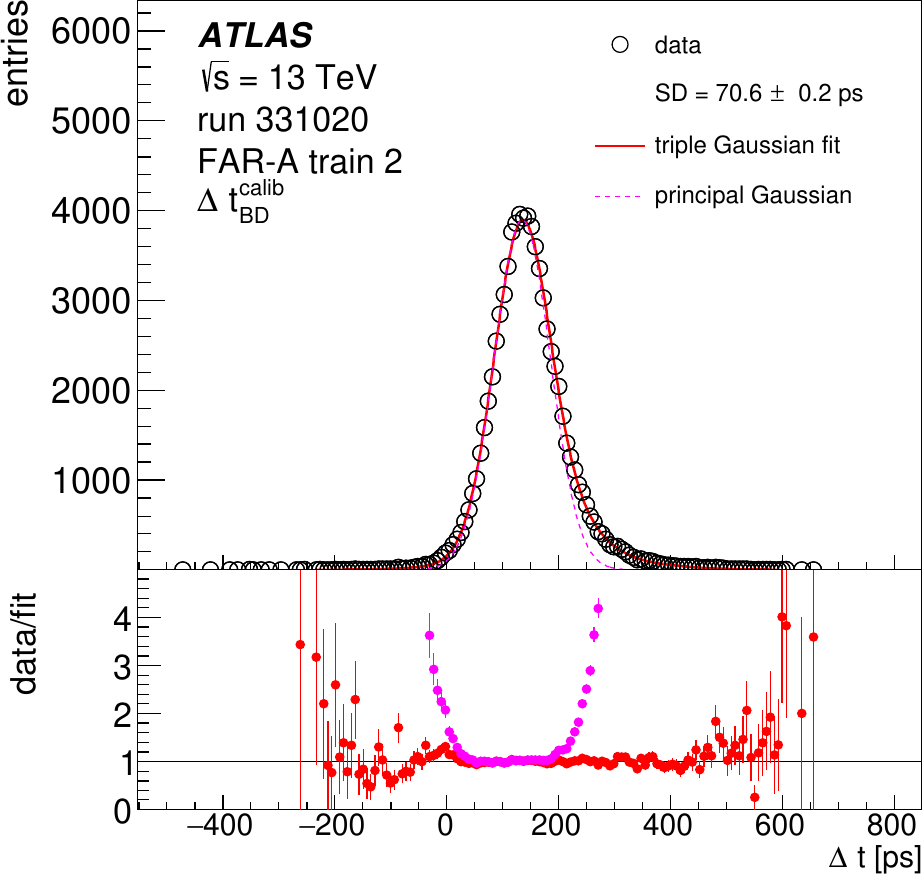}}\hspace{0.3cm}
\subfloat[]{\includegraphics[width=0.46\textwidth]{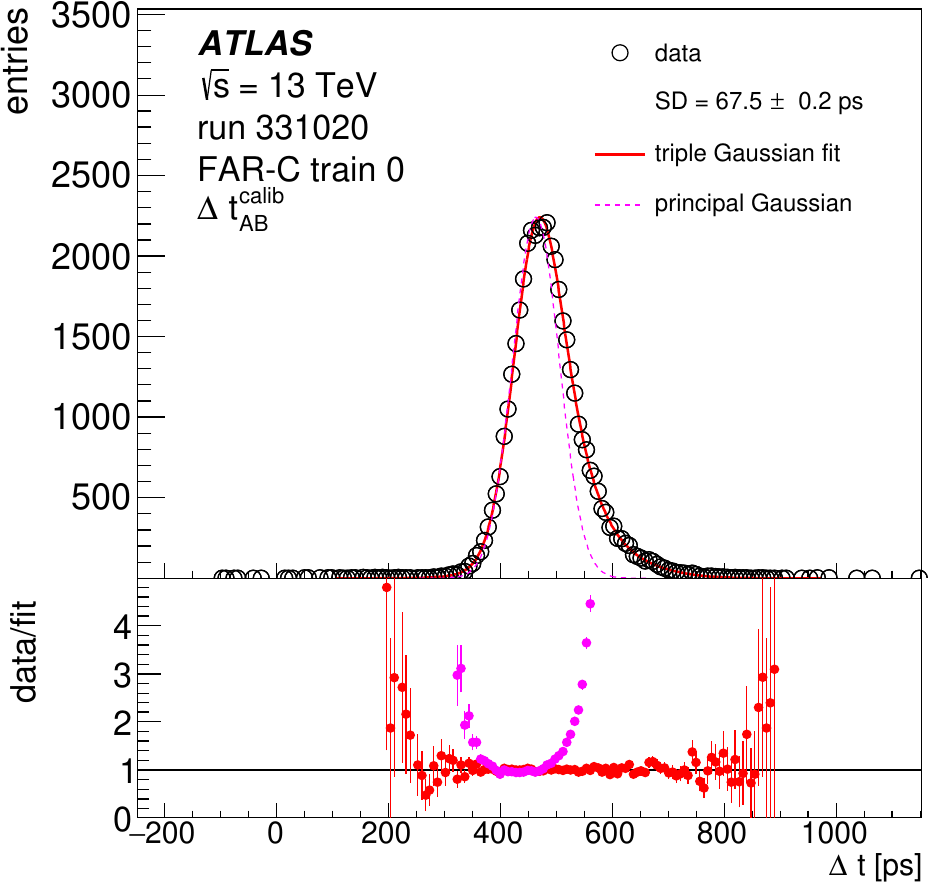}}\\
\subfloat[]{\includegraphics[width=0.46\textwidth]{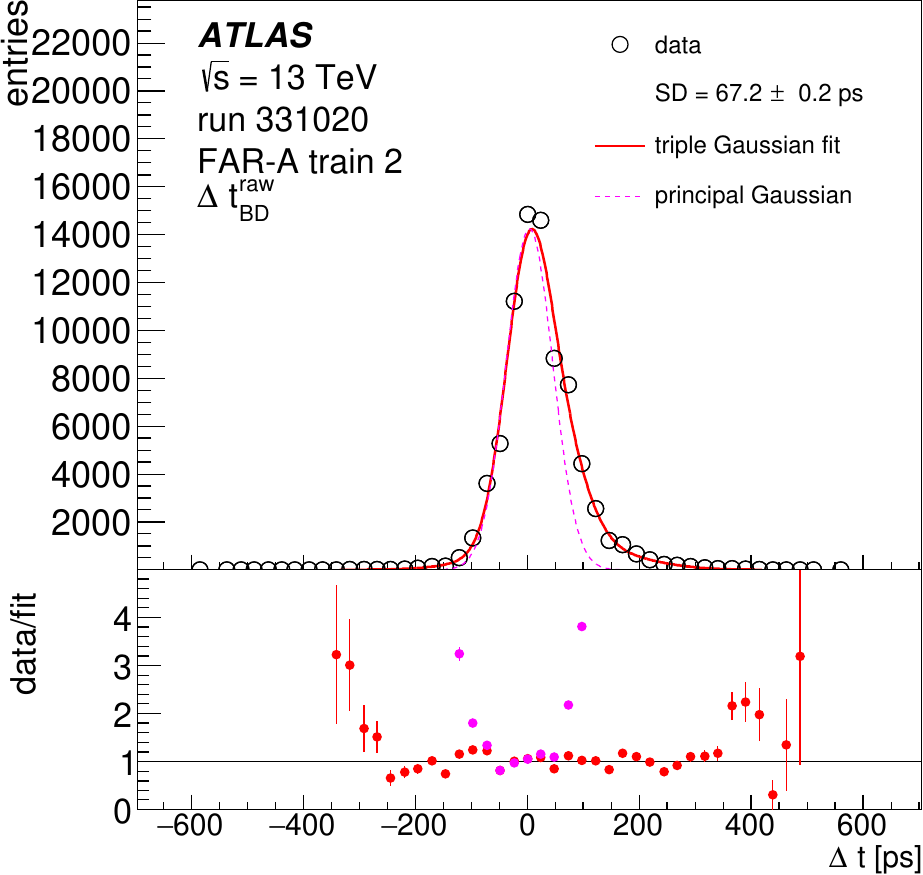}}\hspace{0.3cm}
\subfloat[]{\includegraphics[width=0.46\textwidth]{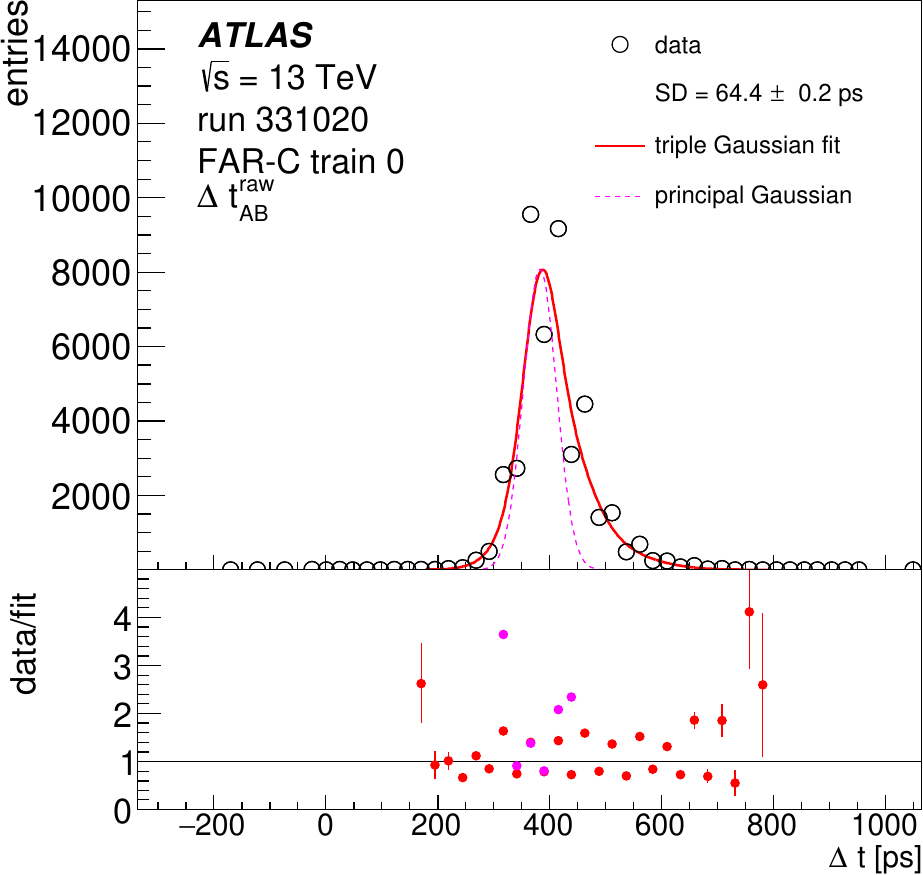}}
\caption{Time-difference distributions measured in selected channel combinations in run 331020. Plots (a) and (b) show the calibrated-time distributions including the bin position correction and uniform smearing within the corresponding HPTDC bins. Plots (c) and (d) show the uncalibrated $\Delta t$ distributions obtained from the raw-time information for the same choice of channels. The data are shown by open circles; the solid line  shows a triple Gaussian fit; the dashed line shows the fit component with maximum integral normalised to the maximum value of the triple Gaussian and is denoted as the principal Gaussian. The data-to-fit ratios are shown in the bottom panels with markers carrying the colour of the above presented fits. The standard deviation of the data is quoted as SD in the legend.}
\label{fig:timediffexamples}
\end{figure}

Similarly as in the efficiency studies, in order to suppress secondary activities in the ToF sensitive volume, the single-ToF-train selection is used as the nominal selection throughout the resolution studies. In Figure~\ref{fig:timediffexamples} examples of the time-difference distributions measured for two selected channel combinations from different trains and stations in run $331020$ are shown. The choice of these channels is made to exemplify different possible shapes of calibrated $\Delta t$ distributions as well as to demonstrate the possible HPTDC binning artefacts in case of using the uncalibrated times. The distributions in the top row correspond to the raw times corrected for the bin centre position and uniformly smeared within the HPTDC bin width (as explained in the previous section and denoted as "calib"), while in the bottom row plots, raw HPTDC channel times are shown. The calibrated results are free of large bin variations introduced by the HPTDC and facilitate the discussion of the $\Delta t$ shapes. Triple Gaussian fits are used to describe the shapes of the calibrated $\Delta t$ distributions satisfactorily. Deviations from a single Gaussian shape are clearly manifested as the extra tails on one or both sides of the distributions. Here, the single Gaussian shape is represented by the component with the highest normalisation of the triple Gaussian model, and is denoted as the principal one. These tails may be attributed to time-walk effects introduced by the variable amplitudes of the input signals not fully compensated by the CFD. The information about the amplitudes, usually approximated by the time over a threshold value, is not available in the data and the time walk effects cannot be directly parameterised and included in the fits.
 
By measuring time differences between individual channels of a single train in one event, dependencies on $t_{\mbox{\tiny proton}}$ and $t_{\mbox{\tiny clock}}$ are eliminated and the following relation is obtained:
\begin{equation}
\Delta t_{ij} = t_{i} - t_{j} = t_{i,\mbox{\tiny delay}} - t_{j,\mbox{\tiny delay}} + t_{i,\mbox{\tiny smear}} - t_{j,\mbox{\tiny smear}}.
\label{eq:channeltimediff}
\end{equation}
Because the time-delay values are constant for each channel, the variance of the time difference distribution follows directly from the variances of the $t_{\mbox{\tiny smear}}$ distributions for the two contributing channels.
 
In the case of four ToF channels per train, six different $\Delta t_{ij}$ channel combinations can be considered: (AB, AC, AD, BC, BD, CD). In general, the $t_{\mbox{\tiny delay}}$ terms cause the $\Delta t_{ij}$ distributions to peak at non-zero values and are treated as nuisance parameters. The widths of the $\Delta t_{ij}$ distributions, $\sigma_{ij}$, enter the extraction procedure of individual channel resolutions, $\sigma_{i}$. It has been found that the second central moments of the triple Gaussian fits coincide well with the standard deviations of the data. The standard deviations are thus used as an input to the extraction of the resolutions. Since the standard deviations depend on the calibration choice only weakly, both the calibrated and raw values are used to extract the resolutions, and their differences are treated as a systematic uncertainty.
The $\sigma_{ij}$ values are related to the resolutions of the corresponding channels as follows:
\begin{equation}
\sigma_{ij}^{2} = \sigma^{2}_{i} + \sigma^{2}_{j} - 2 \rho_{ij} \sigma_{i} \sigma_{j},
\label{eq:sigmaijdiff}
\end{equation}
where $\rho_{ij}$ indicates that the $t_{i,\mbox{\tiny smear}}$ and $t_{j,\mbox{\tiny smear}}$ measured in the same train may not be statistically independent. Such correlations can be caused by charge sharing on the adjacent anode pads of the MCP-PMT leading to modifications of pulse shapes and of the timestamp determination. Another cause of correlated time measurements can be the fact that the MCP-PMT readout electronics share a common ground.
 
To estimate the influence of the correlation factors the data from test beam campaigns were analysed. The MCP-PMTs tested in the 2016 test beam measurements~\cite{ATLAS:2016lsc} were of the same type as those installed for the actual 2017 data taking. The correlation values fall consistently within the range of 0 to 0.2. As a result, selecting $\rho_{ij} = 0$ for all channels is considered the standard choice, while the introduction of an alternative $\rho_{ij} = 0.2$ is taken as generating a systematic uncertainty.
 
The six equations $\sigma_{ij} = \sqrt{\sigma_{i}^{2} + \sigma_{j}^{2} - 2 \rho_{ij} \sigma_{i} \sigma_{j}}$ represent an over-determined system, and the extracted resolution values are those that minimise the following $\chi^{2}$-like expression
 
\begin{equation}
\chi^{2} = \Sigma_{ij} \frac{\left(\sigma_{ij} - \sqrt{\sigma^{2}_{i} + \sigma^{2}_{j}- 2 \rho_{ij} \sigma_{i} \sigma_{j}}\right)^{2}}{\left(\delta_{\mbox{\tiny stat}}(\sigma_{ij})\right)^{2}},
\label{eq:resolutionchi2}
\end{equation}
where $\delta_{\mbox{\tiny stat}}(\sigma_{ij})$ is the statistical uncertainty of $\sigma_{ij}$.

Single-channel resolutions measured for the single-ToF-train hit topologies are presented in Figure~\ref{fig:timeresolutions} for all the studied 2017 runs. The statistical uncertainties obtained from the minimisation procedure using the formula in Equation~(\ref{eq:resolutionchi2}) against the measured $\Delta t$ standard deviations are below $1$~ps. The inactive channels A and D in trains 0 on the A and C sides, respectively, did not contribute to the resolution extraction procedure.
 
A general observation is that channel resolutions in the late 2017 runs are better than those from the earlier runs. The resolutions vary between 20 and 40~ps for the late runs in all channels, except for the A-channels of train~1 and 3 for the FAR-C station. The earlier runs show, on average, worse resolutions, especially for side A, exceptionally reaching values of 70 or 80~ps. These worse resolutions in runs 331020 and 336505 may be caused by a too high voltage applied in the MCP-PMT which led to large output pulses saturating the amplifiers hence giving distorted pulses on the CFD inputs and consequently affecting the precision of the timestamp determination.

It is useful to note that there is a systematic dependence of the resolutions on the channel number. A photon leakage occurs between the LQ-bars downstream of the proton motion, leading to a gradual photon enrichment of the latter bars. This explains the least favourable resolutions observed in the A channels, improving down to the C channels while the last D channels perform similarly to the B channels. The worsening in the D-channels is presumably present due to a lesser charge sharing from neighbouring channels.
 
\FloatBarrier
 
\subsubsection*{Systematic uncertainties}
The following sources of systematic uncertainties are considered in the measurement of the single-channel resolutions:
\begin{itemize}
\item{\textsc Correlation:}
The correlation parameter $\rho_{ij}$ is changed from 0 to 0.2. The resulting alteration in the extracted single-channel resolutions is employed as a measure of systematic uncertainty, and this uncertainty is treated as symmetric.
\item{\textsc Calibration:} $\sigma_{ij}$ are evaluated separately with and without the HPTDC calibration described in section~\ref{hptdc}.
\item{\textsc Event selection:} As an alternative event selection, an additional requirement of a single track reconstructed in the SiT is used, leading to an overall improvement of the resolutions in all channels.
The absolute value of the difference of resolutions for the modified and the nominal event selection is conservatively considered as a symmetric systematic uncertainty.
\end{itemize}
 
\begin{figure}[tbh!]
\centering
\includegraphics[width=1\textwidth]{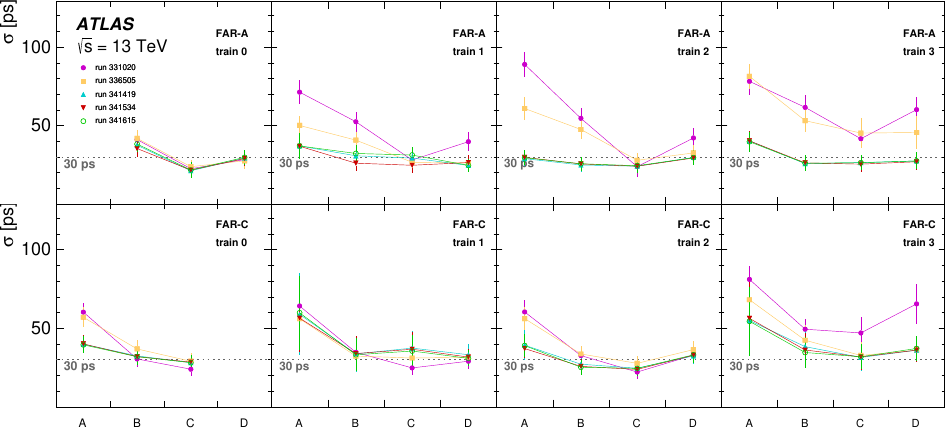}
\caption{Time resolutions measured for individual ToF channels of the AFP station FAR-A (upper row) and FAR-C (lower row) in runs 331020 (full circles), 336505 (full squares), 341419 (up triangles), 341534 (down triangles) and 341615 (open circles). The total error bars indicate the statistical and systematic uncertainties added in quadrature, where the systematic uncertainties dominate. A separate visualisation of the statistical uncertainties is not possible as the sizes for all channels are smaller than the marker sizes. The two malfunctioning channels in train 0 of both stations could not be evaluated and their resolutions are not shown.
}
\label{fig:timeresolutions}
\end{figure}

On average, all three sources of systematic uncertainties contribute at a similar level and are more significant than the statistical uncertainties. In detail, similar systematic uncertainties between 3-5~ps are observed due to correlation and calibration. The uncertainties due to event selection range between 1 and 7~ps with the exception of a few channels, where they reach 10 or 20~ps which corresponds to one third of the actual resolution value. The systematic uncertainties are finally added in quadrature together with the statistical ones to form the total uncertainties represented with total error bars in Figure~\ref{fig:timeresolutions}.

\FloatBarrier
 
\vspace{0.5cm}

\FloatBarrier

\section{Vertex matching analysis}
\label{sec:analysisvtxmatching}
 
In this section the capability of the ToF system to measure the $z$-coordinate of the primary vertex of the $pp \rightarrow pXp$ interactions is investigated.
 
The reconstruction of the production-vertex position using the ToF detector relies on the measurement of the proton arrival times on the A and C sides (double-tag). The proton arrival time depends on the time when the interaction happened, which is not known. It also depends on the position of the interaction. One of the measured arrival times is advanced while the other one retarded proportionally to the vertex position. The difference between the arrival times on both sides does not depend on the interaction time but only on the vertex position, $\ztof$, relative to the nominal IP, i.e.
\begin{equation}
\ztof = \frac{c}{2}(t_{\mbox{{\tiny FAR-C}}}-t_{\mbox{{\tiny FAR-A}}}),
\label{eq:tofformula}
\end{equation}
where it is assumed that the ToF stations are located at equal distances from the origin of the coordinate system and $c$ refers to the speed of light in vacuum.
 
In order to test the $\ztof$ reconstruction, it can be compared to the reconstructed primary vertex $z$-position measured precisely by the ATLAS central detector, $\zatlas$. The primary vertex denotes the vertex with the highest sum of $p_T$ of tracks associated with it and it is expected to be the production point of the hardest $pp$ interaction in the given bunch-crossing. The events where the $pp\rightarrow pXp$ processes are produced in the primary vertex represent the signal combination of $\ztof$ and $\zatlas$. The width of the $\zatlas - \ztof$ distribution reflects a combined resolution of the $\ztof$ and $\zatlas$ measurements. The influence of $\zatlas$ resolution (at the level of $\sim$ 30--100~$\mu$m) can be neglected in the following analysis. If pile-up interactions are present the distribution of $\dz$ is affected by the combinatorial background, see Figure~\ref{fig:processes}. In such events, the width of the $\zatlas - \ztof$ distribution is driven mainly by the longitudinal length of the beam spot~\cite{Cerny:2020rvp}. As discussed earlier, the non-collision background and the genuine leading protons (including the pile-up ones) give the same distributions of arrival times measured by the ToF detectors, leading to the same shapes of the $\dz$ distributions for the pile-up combinatorial and the non-collision backgrounds.
 
\subsection{Data and event selection}
Additional requirements are needed for the vertex-matching analysis. The value of $\mu$ must be as low as practically possible, to minimise the large combinatorial background. The datasets should be of high statistics since the efficiencies of single ToF channels are rather low as reported in section~\ref{sec:resolution}.  For a part of the following analysis,  information about the $\zatlas$ must be available on an event-by-event basis, which precludes the use of the AFP calibration stream.
Three ATLAS runs with average $\mu \sim 2$, labelled as 341419, 341534 and 341615, recorded in November 2017 (see Table~\ref{tab:runs}) fulfilled these requirements. These datasets were recorded with very relaxed trigger conditions denoted commonly as minimum bias.

To maximise the usable statistics, no SiT-related selections are applied. The final event selections were applied to the ToF systems on sides A and C independently, and the ToF signal is required to be in each case in a single train only, with at least one channel hit in the train. The events passing these conditions on one side are called single-tag whereas the double-tag events pass the selection conditions on both sides. The information from two channels with inferior resolution in the side-C station (channels A in trains 1 and 3, see Figure~\ref{fig:timeresolutions}) is excluded.
 
\subsection{Determination of channel time delays}
\label{sec:analysisvtxmatchingtimedelays}
 
The time measured by a single ToF train is obtained as an average of the times measured in its individual channels. To do this, the time delays of the channels have to be taken into account.
The $\tdelay$ values (see Equation~\ref{eq:channeltime}) can be in principle constrained by studying the mean values of $\Delta t_{ij}$ distributions between channels of a given train as can be seen from Equation~\ref{eq:channeltimediff}. In this analysis another approach has been adopted.
 
The changes of mean $z$-position of the ATLAS beamspot, $\zbs$, translate linearly to changes in the arrival times measured in the ToF detectors, where the $\zbs$ information is updated for each lumiblock. Any combination of two hit channels, one in the FAR-A, the other in the FAR-C station provides a prediction for the $\ztof$ observable given by Equation~\ref{eq:tofformula} (see e.g. Figure~3 in Ref.~\cite{Cerny:2020rvp}). The $\ztof$ value can also be obtained using two unrelated single-tag events (mixed events, ME) according to Equation~\ref{eq:tofformula} and will be referred to as $\ztofme$. The random pairs of single-tag events form a multitude of quasi-double-tag combinations whose mean value of $\ztofme$ carries the information about the $\zbs$. For each lumiblock, the event mixing is performed using only events from this particular lumiblock.

The event mixing procedure described above leads to a numerically improved distribution of statistically equivalent values of $\ztofme$. For example using ten values of time measured in the FAR-A side and ten in the FAR-C side, recorded in separate events in a given lumiblock can ideally lead to 100 $\ztofme$ hypotheses. Note that no genuine double-tag measurement is explicitly required. The actual mixing is done randomly with the maximum number of created combinations limited, such that the probability of re-using any combination more than once is kept below $0.001$.
The sizes of the mixed-event samples are typically a few hundreds up to thousand of mixed events per lumiblock for each single choice of A- and C-side channels.
 
To address the influence of channel delays the following relation is used:
 
\begin{equation}
\langle \ztof^{(ij)\,\mbox{\tiny ME}}\rangle + s_{A}^{i} -s_{C}^{j} = z_{\mbox{\tiny BS}}, \nonumber
\end{equation}
 
where the $i$ and $j$ indices refer to the chosen channel in the FAR-A and FAR-C station, respectively, the ME superscript refers to the event-mixing method, the $s_{A}^{i}$ and $s_{C}^{j}$ terms denote corrections of the  $\tdelay$ values of the channels which are treated as constants and expressed in millimetres. The delays are channel-specific constants which for each $ij$-pair are known up to an additive constant. Only their differences matter, which means that one of the delay values in the pair can be chosen arbitrarily. Because there are $16\times16$ pair combinations, each channel participates multiple times. In the end, only one channel is then left whose delay can be fixed arbitrarily. Without loss of generality, such a condition is applied to channel~C of train~1 of the FAR-A station, i.e $s_{A}^{\mbox{\tiny train} 1, \mbox{\tiny C}} = 0$~mm. All the channel delays are then determined by obtaining their optimum values based on finding a minimum value of
 
\begin{equation}
\chi^{2} = \sum_{ij} \left[  \frac{ \left\langle \sum_{\mbox{\tiny LB}} \left( z_{\mbox{\tiny BS}}^{\mbox{\tiny LB}} -  z_{\mbox{\tiny ToF}}^{(ij)\,\mbox{\tiny ME,LB}}\right)\right\rangle_{\mbox{\tiny fit}} - s_{A}^{i}+ s_{C}^{j}  }{\delta^{(ij)}_{\mbox{\tiny fit}}}\right]^{2},
\label{eq:ztofACeq2}
\end{equation}
 
where the outer sum runs over all channel pairs $ij$ and the fitted mean value for each $ij$ pair is obtained from a Gaussian fit to a distribution constructed as a sum of mixed-event distributions from all considered lumiblocks. The statistical uncertainties on the mean values fitted in each channel combination are denoted as $\delta_{\mbox{\tiny fit}}$.
 
The extracted corrections make the numerators in the sum in Equation~\ref{eq:ztofACeq2} as small as $0.1-0.2$~mm for each fixed combination of $(ij)$. Only in runs 341534 and 341615 the uncertainty grows significantly in the channel combinations where FAR-A train 3 channel A participates, and it was decided to mask this channel in the subsequent analysis.
 
\subsection{Closure test of channel delays and comparison with beam-spot}
\label{sec:closure}
\begin{figure}[t!]
\centering
\subfloat[]{\includegraphics[width=0.32\textwidth]{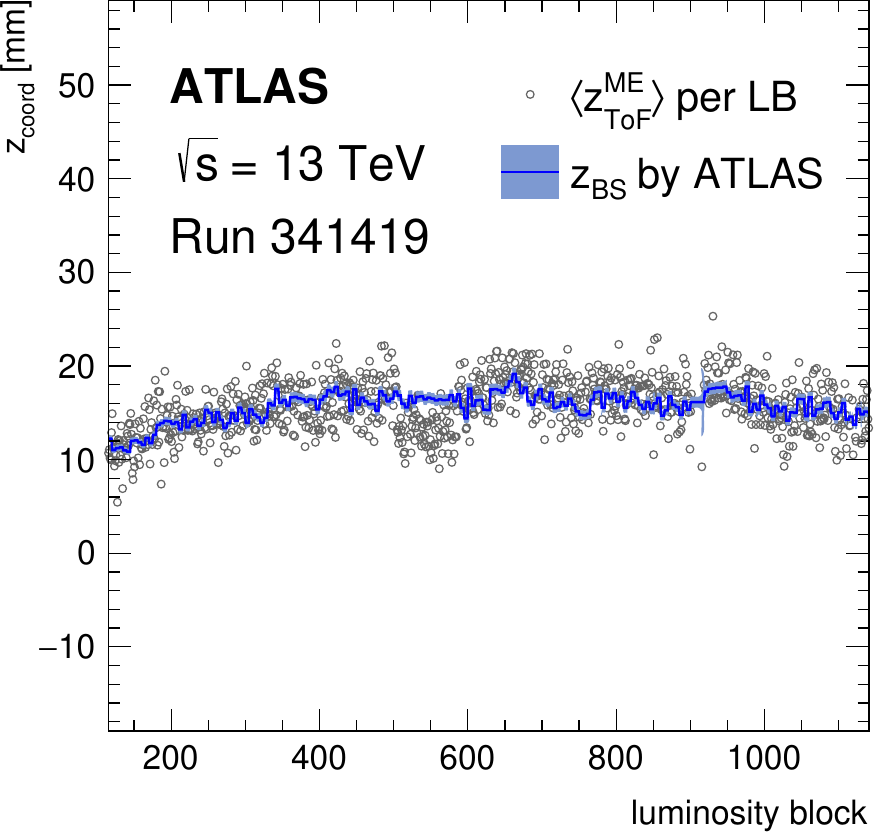}}\hspace{0.15cm}
\subfloat[]{\includegraphics[width=0.33\textwidth]{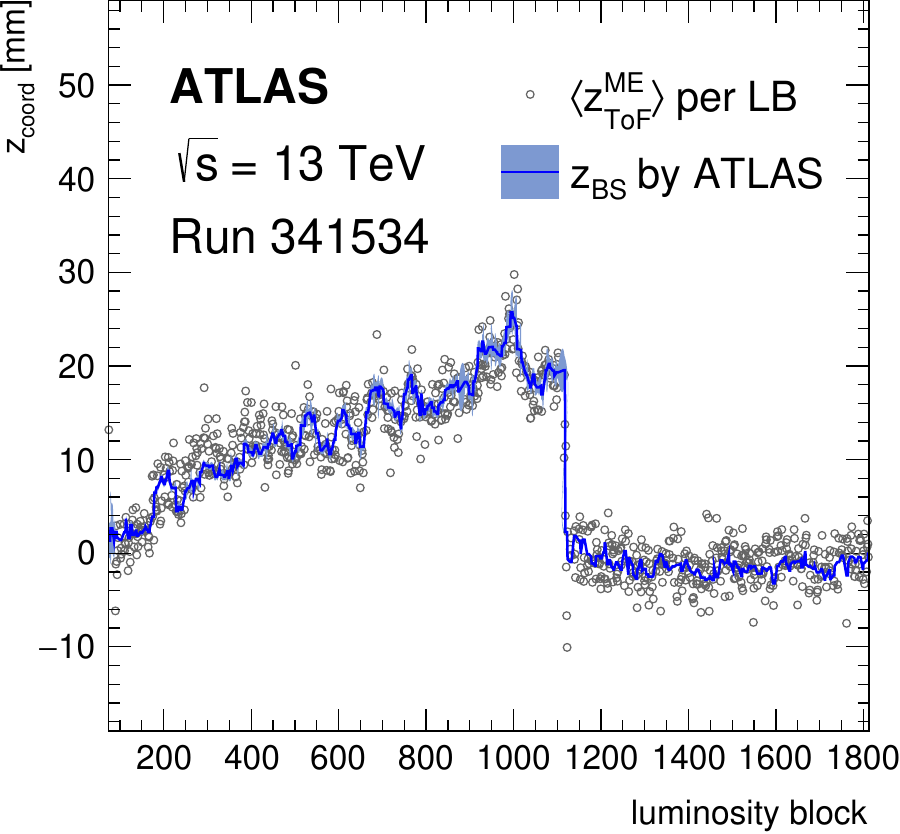}}\hspace{0.15cm}
\subfloat[]{\includegraphics[width=0.32\textwidth]{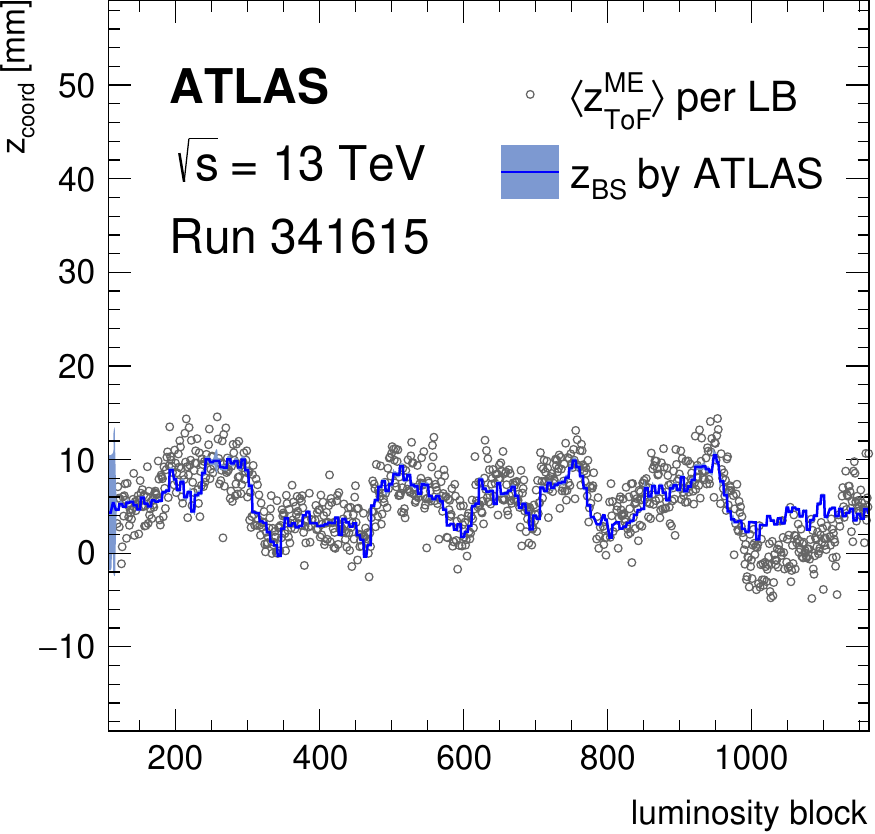}}\\
\subfloat[]{\includegraphics[width=0.32\textwidth]{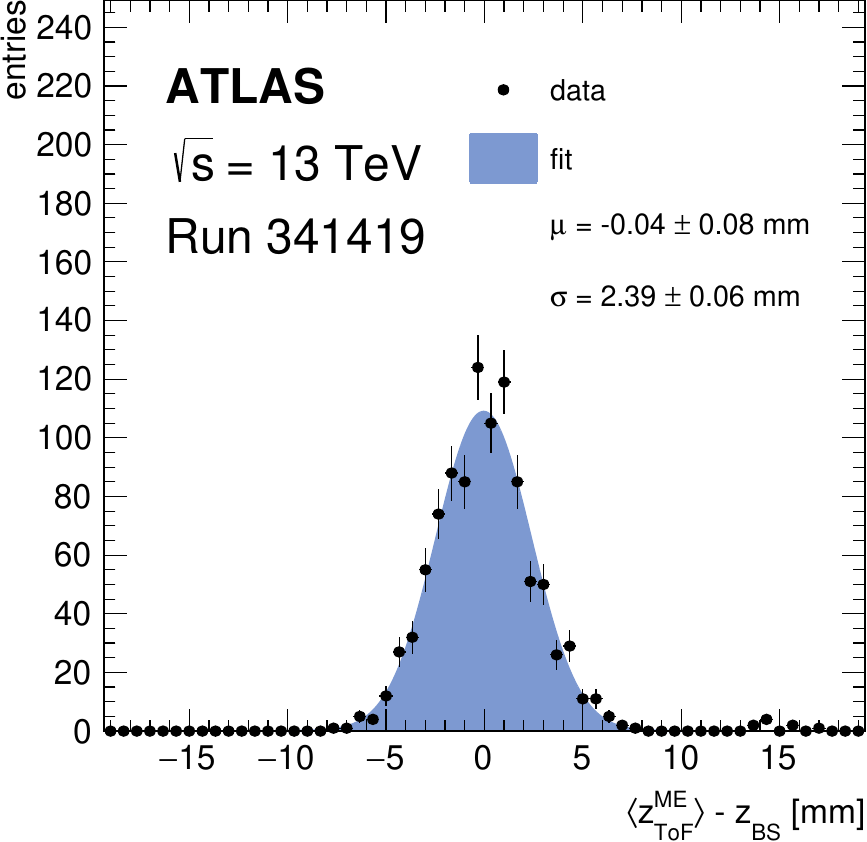}}\hspace{0.15cm}
\subfloat[]{\includegraphics[width=0.32\textwidth]{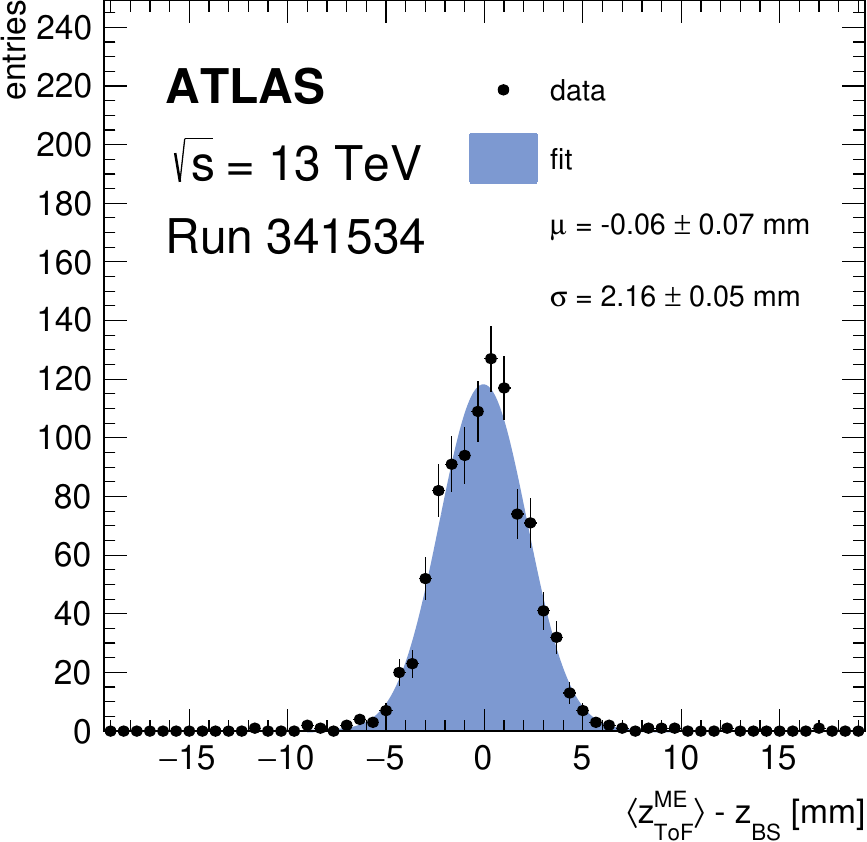}}\hspace{0.15cm}
\subfloat[]{\includegraphics[width=0.32\textwidth]{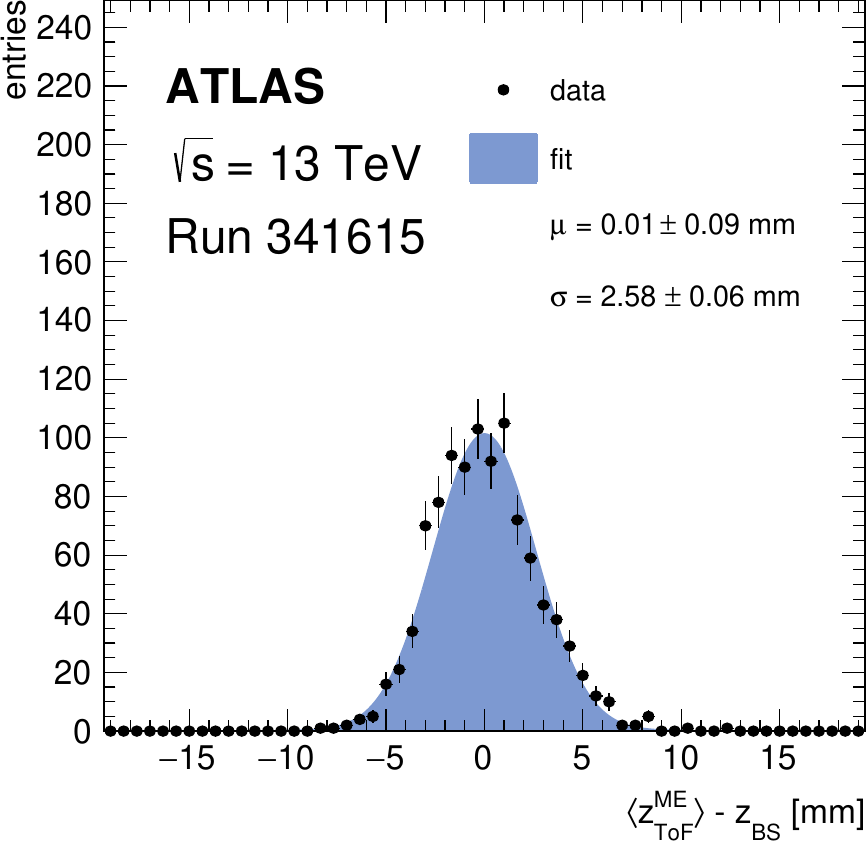}}
\caption{Panels (a)--(c) show the mean values of Gaussian fits to the mixed-event distributions of $z_{\mbox{\tiny ToF}}^{\mbox{\tiny ME}}$ in each lumiblock (open circles with no uncertainty indicated) compared with beam-spot $z$-coordinates $z_{\mbox{\tiny BS}}$ measured by the ATLAS central detector shown by the histogram where the band indicates its uncertainty. Panels (d)-(f) show the distribution of the difference $\langle z_{\mbox{\tiny ToF}}^{\mbox{\tiny ME}}\rangle - z_{\mbox{\tiny BS}}$ from all lumiblocks (points) compared with a Gaussian fit (filled area).  The vertical axis label `entries' represents the number of lumiblocks. The plots from left to right column correspond to runs 341419, 341534 and 341615.}
\label{fig:zBSclosure}
\end{figure}

Figure~\ref{fig:zBSclosure} represents a closure test of the delay corrections extracted by the procedure described above for all three considered runs. The plots in the top row compare the mean values of Gaussian fits to the $\ztof$ distributions from mixed-event samples, $\langle\zbs^{\mbox{\tiny ME}}\rangle$, with the $z$-coordinate of the beam-spots measured by the central ATLAS detector, $\zbsatlas$, for each lumiblock considered. Although the lumiblock evolution of the $\zbs$ quantity is strongly run-specific, it is well reproduced by the ToF, apart from two short periods, one in run 341419 and one in run 341615. The reason for these small non-closures was not identified. A more detailed consistency check, i.e. the distribution of $\zbsatlas - \langle\zbs^{\mbox{\tiny ME}}\rangle$ from all lumiblocks, is shown in the bottom panels of Figure~\ref{fig:zBSclosure}. The distributions are well described with Gaussian fits centred around zero with widths of about $2.4, 2.2$ and $2.6$~mm. These widths are constituted primarily by the statistical uncertainty of the $\ztof$ measurement originating from the finite size of data in a single luminosity block.
 
\subsection{Expected time resolution}
 
The expected time resolution of the ToF detector can be calculated with the use of the single-channel resolutions measured in Section~\ref{sec:resolution}. In general, the best time resolution is achievable with all four channels of a given train fired. In practice, the hit multiplicity distributions in different trains vary significantly, which is taken into account by an averaging procedure. The correlations between contributing channels are handled in a manner akin to Equation~\ref{eq:sigmaijdiff}, but this time, with a positive sign applied to the cross-terms of the channel pairs. The standard choice of $\rho_{ij} = 0$ for all pairs of channels reflects the scenario of uncorrelated time measurements within the ToF channels. A global variation by 0.2 is taken as a systematic uncertainty whose impact is then taken symmetrically in both directions. The remaining uncertainties of the single-channel resolutions are propagated independently, without correlation.
 
\begin{figure}[tbh!]
\centering
\includegraphics[width=0.8\textwidth]{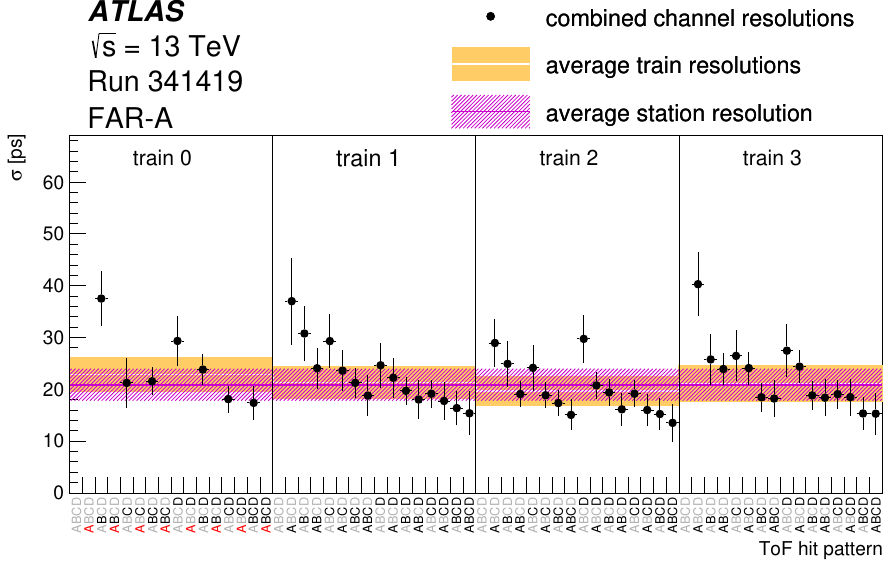}
\caption{The expected time resolutions from single-ToF-train events tagged independently in the FAR-A station in the minimum bias data of run 341419: for the individual hit combinations (points), for all hit combinations measured in single trains (filled bands) and for all single-ToF-train events measured in the stations (hatched band). The bands indicate the uncertainties obtained by propagating quadratic sums of the statistical and systematic uncertainties of single channel resolutions as well as of the statistical uncertainty of the number of each of the hit combinations. The channels that contribute to each given combination are marked by black letters, those which do not by gray letters and finally the inactive channels by red letters.}
\label{fig:avgresolutions341419}
\end{figure}

In Figure~\ref{fig:avgresolutions341419} the expected resolutions are shown by black points for all train multiplicity combinations (denoted as ToF hit pattern) observed in the minimum bias data sample measured by ATLAS in run $341419$ under the condition of single-ToF-train topologies in single-tag events. These timing resolutions are calculated as quadratic sums of the single-channel resolutions divided by the actual train hit multiplicities. Some of the multiplicity combinations are not shown or are not considered in the averaging due to the channel malfunctioning or its exclusion from the analysis (in other words, the channel time information is not used, but the event is kept) discussed in previous sections. As the number of contributing channels increases the resolutions improve. The dominant contribution to the systematic uncertainty is coming from the correlation between channels, and it naturally plays a larger role for events with larger hit multiplicities. However, due to the use of $\rho_{ij} = 0.2$ to account for the systematic uncertainty on the correlations between the channels, the resulting uncertainty may become substantial, as can be observed in the highest multiplicity cases. The train resolutions (shown with shaded bands) are calculated from weighted quadratic sums of resolutions of each multiplicity combination within the train, where the weights are calculated from the actual distributions of multiplicity combinations in the given run. Finally, the train resolutions are combined into the expected station resolutions in the same weighted manner (shown with the hatched band). The uncertainties associated with the resolutions of each multiplicity combination are propagated. Additionally, the statistical uncertainties of the multiplicity distributions are also taken into account for both train and station resolutions.
 
\vspace{0.5cm}
 
In Table~\ref{tab:expresolutions} the expected timing resolutions are presented for the selected runs. The individual A- and C-side ToF timing resolutions are obtained from the respective single-tag data sets as described above. The sufficiently large statistics of these single-tag samples make the statistical uncertainty negligible and the dominant uncertainty comes from propagating the total resolution uncertainty of individual channels. In the last column of the table the single-side resolutions are combined to the expected resolution  according to Equation~\ref{eq:tofformula}.
\FloatBarrier
\begin{table}[hhh!]
\caption{The expected individual-side (second and third column) and combined (last column) ToF resolutions measured in single-tag samples of the minimum bias data in the selected ATLAS runs.}
\begin{center}
\renewcommand{\arraystretch}{1.2} 
\begin{tabular}{|c|c|c|c|}
\multicolumn{1}{c}{Run} & \multicolumn{1}{c}{FAR-A resolution [ps]} & \multicolumn{1}{c}{FAR-C resolution [ps]} & \multicolumn{1}{c}{Combined resolution [mm]}\\
\hline
341419  &  $21\pm 3\,(\mbox{stat}\oplus\mbox{syst})$ & $28 \pm 4\,(\mbox{stat}\oplus\mbox{syst})$ & $5.3 \pm 0.6\,(\mbox{stat}\oplus\mbox{syst})$ \\
341534  &  $20\pm 3\,(\mbox{stat}\oplus\mbox{syst})$ & $28 \pm 4\,(\mbox{stat}\oplus\mbox{syst})$ & $5.2 \pm 0.6\,(\mbox{stat}\oplus\mbox{syst})$ \\
341615  &  $22\pm 3\,(\mbox{stat}\oplus\mbox{syst})$ & $28 \pm 4\,(\mbox{stat}\oplus\mbox{syst})$ & $5.3 \pm 0.6\,(\mbox{stat}\oplus\mbox{syst})$ \\
\hline
\end{tabular}
\end{center}
\label{tab:expresolutions}
\end{table}
\FloatBarrier
 
\subsection{Comparison with the primary vertex position}
In Section~\ref{sec:closure}, a good consistency of the ToF detector measurements was demonstrated at the level of lumiblock and the $\zbs$ observable. In this section it is investigated if a similar consistency is observed when making a direct comparison of the $\zvtxkarel$ measurements on an event-by event-basis by analysing the shape of the $\zatlas - \ztof$ distribution. For this purpose, events with the single-ToF-train signature, observed simultaneously on both the A and C sides are selected. The proton arrival time for a given station and a given event is obtained by averaging over the number of hits, all corrected for the HPTDC bin centres and for the channel delays.
 
\begin{figure}[t]
\centering
\includegraphics[width=0.5\textwidth]{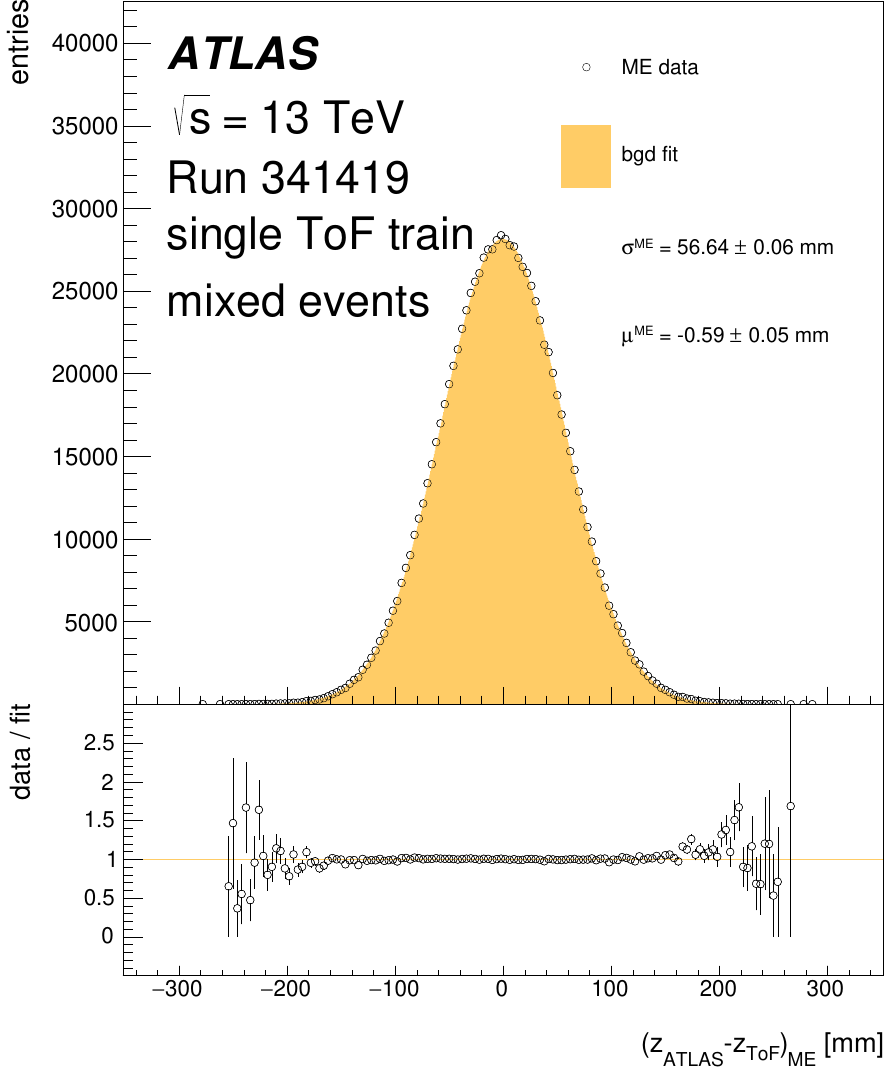}
\caption{The $\zatlas-\ztof$ distribution from the mixed-event data (open circles) used to model the combinatorial background in run 341419. The two-Gausian fit is shown by the filled area. The ratio of mixed data to the fit is shown in the bottom panel of the plot.}
\label{fig:MEfit341419}
\end{figure}

The $\zatlas-\ztof$ distribution for double-tag events has two components: the signal component corresponding to genuine $pp\rightarrow pXp$ processes which is expected to manifest itself as a narrow peak above a broad hump-type distribution corresponding to the second component, the combinatorial background. The latter is expected to follow a double-Gaussian distribution with a special constraint that the width of the secondary component is fixed to the width of the main component divided by $\sqrt{2}$, which stems from different configurations contributing to the combinatorial background, as indicated in Figure~\ref{fig:processes} or see~\cite{Cerny:2020rvp}.
The shape parameters of the background component are obtained from a fit to the mixed-event sample prepared from combinations of unrelated values of the FAR-A, FAR-C side times and the $\zatlas$ values from all the lumiblocks now (using the similar reasoning as in section~\ref{sec:analysisvtxmatchingtimedelays}). The values that are combined correspond to average train times obtained under the single-ToF-train condition on the given side. The mixing of single-tag events gives more statistics for a better background shape estimation while remaining independent of the double-tag events which are the primary focus of this vertex-matching analysis. Figure~\ref{fig:MEfit341419} shows the $\zatlas-\ztof$ distribution from the mixed-event sample together with a double-Gaussian fit and the extracted parameter values in run 341419. The fitted mean value is again reasonably close to zero given the widths obtained in the closure tests documented in Figure~\ref{fig:zBSclosure} bottom plots. These parameters are then used in the subsequent vertex-matching analysis. In this run the secondary Gaussian component is found negligible.
 
\begin{figure}[tb!]
\centering
\subfloat[]{\includegraphics[width=0.48\textwidth]{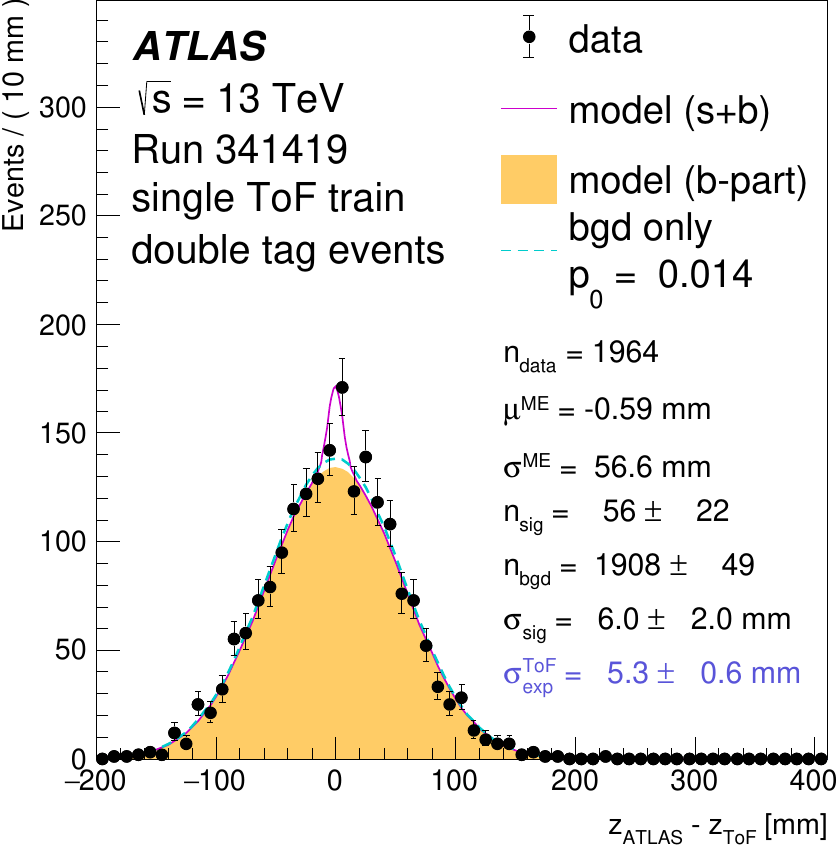}}\hspace{0.2cm}
\subfloat[]{\includegraphics[width=0.48\textwidth]{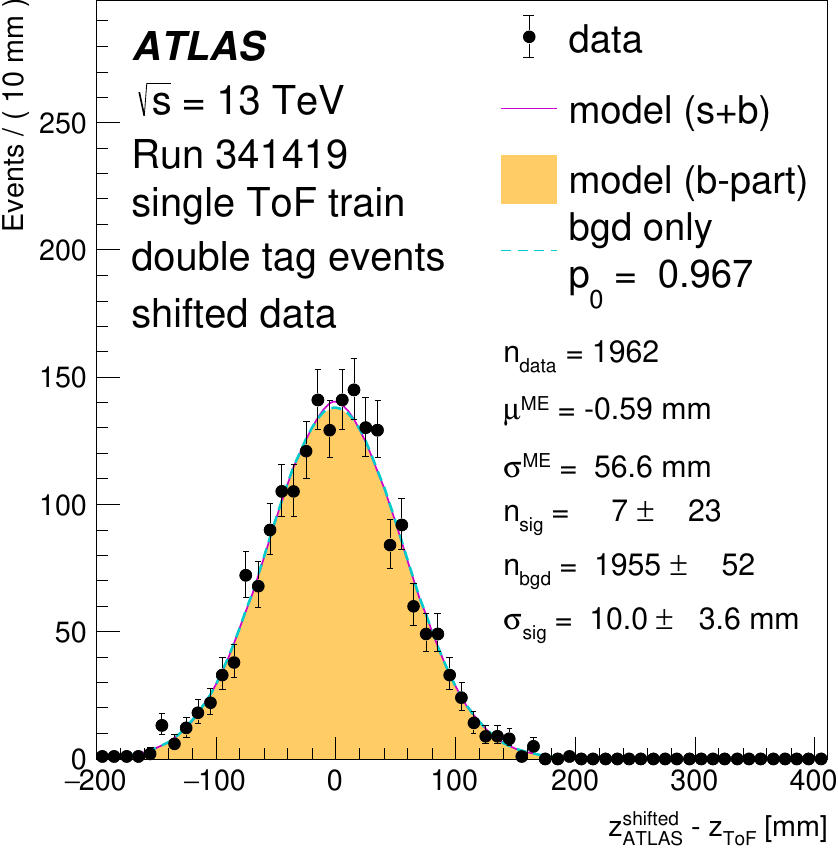}}
\caption{The $\zatlas-\ztof$ distributions for the nominal data in (a) and $\zatlas^{\mbox{\tiny shifted}}-\ztof$ for the shifted data in (b) measured in run 341419 (dots) using double-tagged events with the single-ToF-train condition. The fits of the sum of signal and background components are represented by solid lines. The background component is indicated by the filled area. The background-only model fits are shown by the dashed lines. In the legends the values of fitted parameters, $\nsig, \nbgd$ and $\sigmasig$ are quoted together with the p-values ($p_{0}$) corresponding to the background-only hypothesis.}
\label{fig:deltazfit341419alltr}
\end{figure}

The $\dz$ distribution is now analysed to extract  information about the $\zvtxkarel$ resolution for the signal events. The results presented here are taken from double-tagged events recorded in the large-statistics low-$\mu$ run 341419 with ToF channel times calibrated for HPTDC bin centers and fully determined delays. The left panel of Figure~\ref{fig:deltazfit341419alltr} shows the actual $\zatlas-\ztof$ distribution while the right panel shows the $\zatlas^{\mbox{\tiny shifted}}-\ztof$, where $\zatlas^{\mbox{\tiny shifted}}$ is taken from the next event, thus, by definition unrelated to $\ztof$ from the signal, for clarity denoted also as `shifted data' in the legend or `shifted' in the following paragraphs. The parameters of the signal and background mean and the background shape are fixed from the previous fit to the ME data (see Figure~\ref{fig:MEfit341419}). The remaining free parameters are thus the numbers of signal and background events and the signal width, $\nsig, \nbgd$ and $\sigmasig$, respectively. Unbinned fits are performed, using the extended likelihood method as implemented in RooFit~\cite{verkerke2003roofit}.

The $p$-value for the background-only hypothesis, $p_{0}$, equal to 0.014 for the nominal data, which corresponds to a significance of 2.2~$\sigma$, suggests that the distribution is consistent with the presence of a signal, as shown in  Figure~\ref{fig:deltazfit341419alltr}~(a). This is also supported by the fit itself yielding a significant fraction of signal events in the total studied sample and the width, $\sigmasig$, of 6.0\,$\pm$\,2.0~mm, consistent with the expected value (5.3\,$\pm$\,0.6~mm) obtained from the single-channel resolution values in the double-tagged event sample. The right panel then shows that the shifted data where the signal $\dz$ combinations cannot be present are indeed compatible with the background-only hypothesis, as seen from a high $p_{0}$ value of 0.951. This observation indicates that the ToF detector operates as intended and allows for the measurement of the vertex position in double-tag events.
 
\begin{figure}[tbh!]
\centering
\subfloat[]{\includegraphics[width=0.48\textwidth]{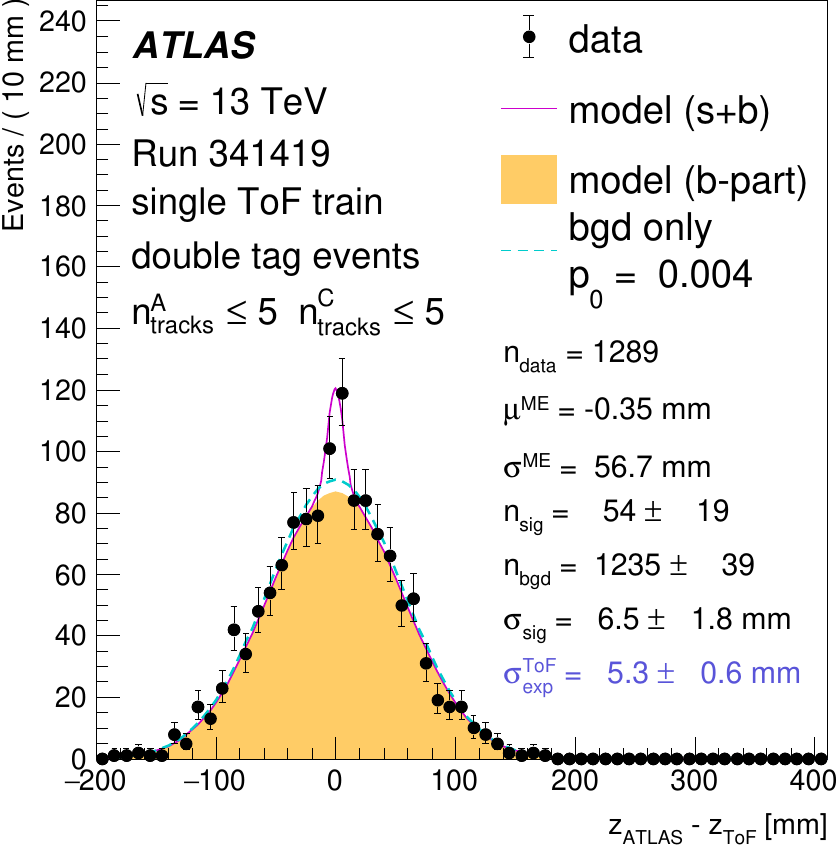}}\hspace{0.2cm}
\subfloat[]{\includegraphics[width=0.48\textwidth]{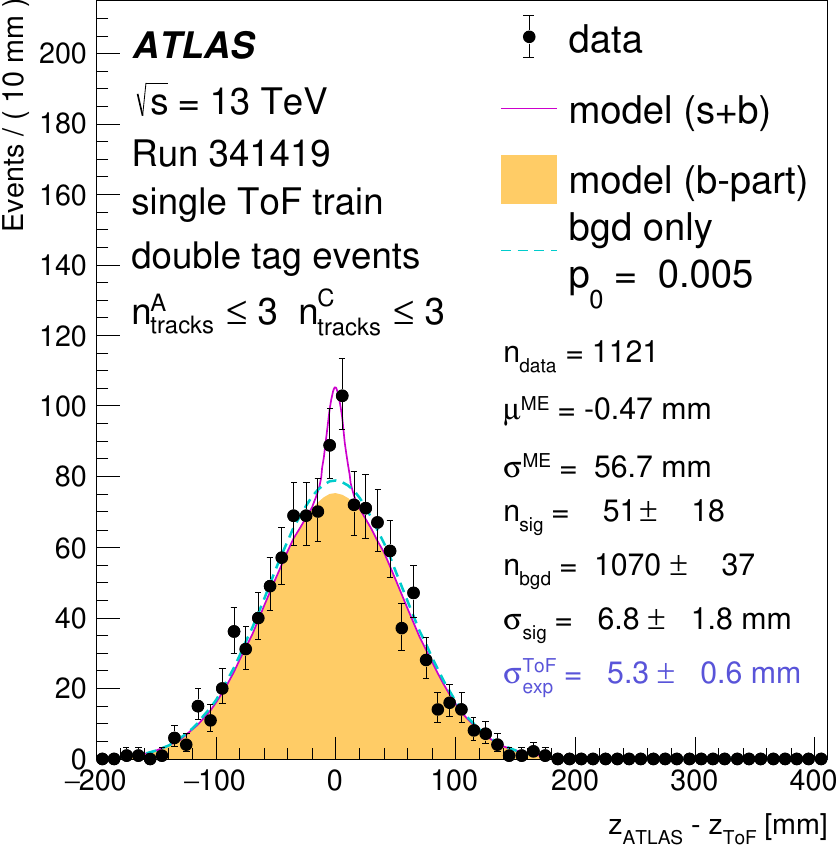}}\\
\subfloat[]{\includegraphics[width=0.48\textwidth]{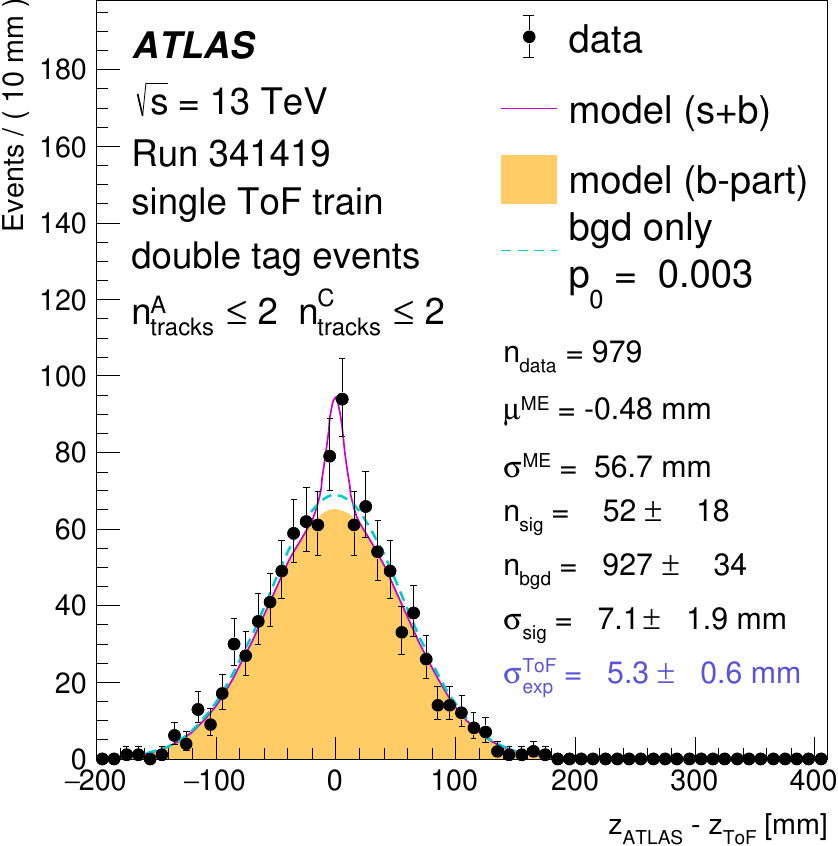}}\hspace{0.2cm}
\subfloat[]{\includegraphics[width=0.48\textwidth]{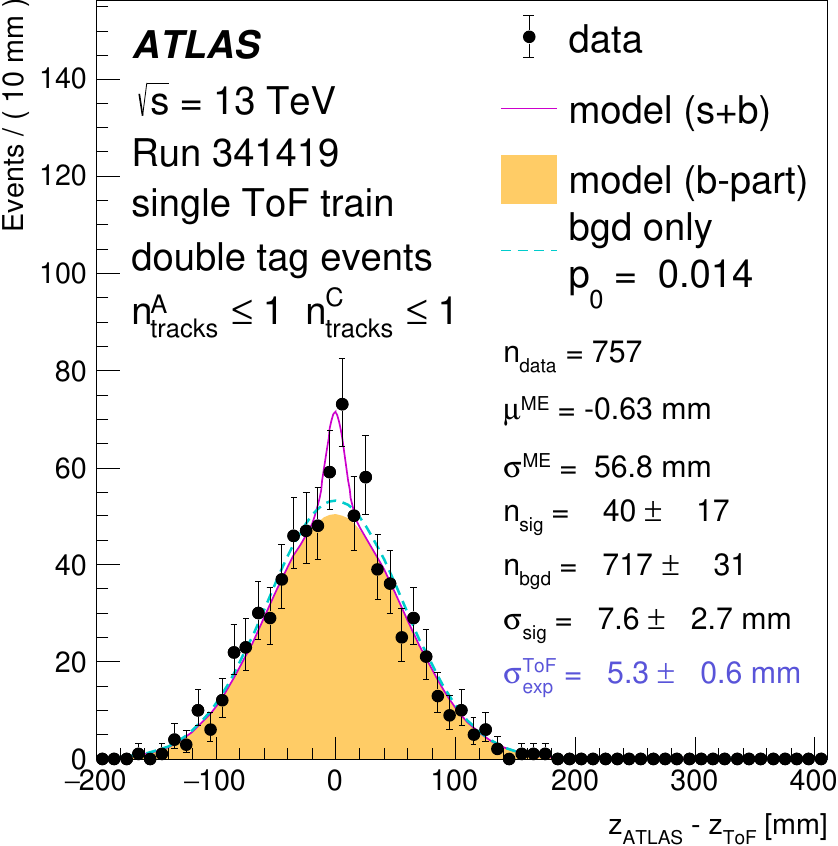}}
\caption{From (a) to (d) the $\zatlas-\ztof$ distribution and the fit results obtained in run 341419 corresponding to tightening the cut on a maximum number of reconstructed AFP SiT tracks pointing to the relevant station.}
\label{fig:deltazfit341419trcuts}
\end{figure}

As explained above, the background stems from coincident double-tag detections of protons from pile-up interactions (dominantly from soft SD processes) or from the non-collision background.
It should be noted that no explicit cuts have been applied on the SiT signals. As mentioned in Section~\ref{sec:BIB} the non-collision background contributes less in the events with low SiT track multiplicities. It is therefore useful to check the dependence of the background fraction on such a criterion. Figure~\ref{fig:deltazfit341419trcuts} shows results of such a study, varying the number of SiT tracks and repeating the whole procedure leading to Figure~\ref{fig:deltazfit341419alltr}. It can be seen that tightening the track multiplicity cut to 5, 3, 2 and 1 on both AFP sides does not decrease the signal contribution appreciably, while the background component falls faster. With slightly increasing statistical uncertainty, the extracted resolution is still well compatible with the expected one for all track multiplicity choices.

The analysis of runs 341534 and 341615 proceeds in the same manner as for run 341419. However here, the presence of the signal indicated by the difference of $p_{0}$ values for the nominal and shifted $\dz$ data is not fully supported by values of $\sigmasig$ and $\nsig$ obtained from the fit as documented in Table~\ref{tab:p0values}. The fits tend to converge to narrow signal widths, caused probably by data fluctuations in the expected signal region. Various tests including the use of HPTDC and delay corrections from run 341419 or tighter SiT track-multiplicity cuts with the aim to suppress the background have not helped to promote the signal in runs 341354 and 341615.
In the single-channel analysis using data with the single-tag condition, efficiencies and resolutions were found to be compatible within uncertainties for these runs (see Sections~\ref{sec:efficiency},~\ref{sec:resolution}). Also the delay corrections were found to provide an identical level of precision in describing the $\zbs$ position as documented in Figure~\ref{fig:zBSclosure}. No reason for a different behaviour in these runs has been identified. One possible hypothesis is a worsening of synchronisation between the two ToF detectors on the A and C sides.
 
\begin{table}[tbh!]
\caption{Summary table of values of signal parameters $\sigma_{\mbox{\tiny sig}}$ and $n_{\mbox{\tiny sig}}$ obtained from fits to the nominal $\dz$ data (second and third column) and the background only hypothesis p-values, $p_{0}$, for the nominal and shifted data (last two columns). The values are shown for all the minimum bias data in the selected ATLAS runs.}
\begin{center}
\renewcommand{\arraystretch}{1.3} 
\begin{tabular}{|c|c|c|c|c|c|}
\multicolumn{1}{c}{Run} & \multicolumn{1}{c}{$\sigma_{\mbox{\tiny sig}}$~[mm]} & \multicolumn{1}{c}{$n_{\mbox{\tiny sig}}$} & \multicolumn{1}{c}{Nominal data $p_{0}$} & \multicolumn{1}{c}{Shifted data $p_{0}$}\\
 
\hline
341419  &  $6.0\pm2.0$&$57\pm22$&0.014&0.951\\
341534  &  $4.0\pm1.2$&$41\pm16$&0.013&0.702\\
341615  &  $3.0\pm6.9$&$32\pm14$&0.044&0.373\\
\hline
\end{tabular}
\end{center}
\label{tab:p0values}
\end{table}
 
The effects associated with finite HPTDC widths and the precision of determining channel delays were evaluated using randomly smeared data replicas. Two smearing procedures were performed simultaneously. For each replica, the $\Delta z$ distributions were subjected to fits, as in the nominal case. The measured channel times were uniformly smeared across the actual HPTDC bin width for every event in each replica. Channel delay corrections were modified using a Gaussian distribution centred around the nominal delay value, with a conservative width corresponding to 2.5~mm$/c$. This choice is inspired by the values from Figure~\ref{fig:zBSclosure}, where the individual delay alterations remained fixed for each data replica. The fits, applied repeatedly to each of the unique data replicas, result in model parameter values and associated statistical uncertainties that are no longer fixed. The mean value of the $\sigmasig$ and its statistical error yield $\sigmasig = 6.0 \pm 2.1$ mm. This implies a negligible difference compared to values obtained from the fit to the original data. Consequently, it can be concluded that the impact of these two sources of ambiguity can be disregarded.

\FloatBarrier
\section{Conclusions}
 
The performance of the AFP Time-of-Flight detector system in ATLAS has been measured using data collected in 2017 at the LHC. Efficiencies of ToF trains and timing resolutions of individual ToF channels are obtained from single-tag events in high-statistics AFP calibration-stream runs, while double-tag events in the minimum-bias runs are used to study the compatibility of the measurement of the $z$-coordinate of the primary vertex by the ToF detector with that of the central ATLAS tracker. The methods and approaches used in this performance study are largely universal, and are hence applicable regardless of the specific detection techniques employed for measuring arrival times in the forward region. This makes them potentially useful as guidelines for the effective utilisation of future data from the Time-of-Flight detector system.
 
The efficiencies of individual trains are measured for two scenarios which are sensitive to possible contributions of secondary particles. The baseline requirement for both is exactly one local track reconstructed in the Silicon tracker in a given AFP FAR station. For the early data used in this study, the train efficiencies are measured at percent levels reaching up to 20\%, degrading with time in 2017 to sub-percent levels for the last studied runs. The rapid decrease of the efficiencies is consistent with an exceeded MCP-PMT lifetime. A clearly visible fraction of events containing secondaries from particle showers is well suppressed by requiring in addition that only a single train is seen in the whole sensitive ToF area. This selection reduces event yields by about a factor of two, but provides favourable conditions for the timing resolution measurements.
 
The resolutions of individual ToF channels were extracted from the analysis of time differences between the channels of a single train where the events were required to contain hits in just the studied train. Typical single-channel resolutions vary between 23 and 40~ps for three late 2017 runs, and between 30 and 80~ps for the two earlier ones, where the worse resolutions can be explained by a saturation of preamplifiers caused by too high a gain of the MCP-PMTs. The measured single-channel resolutions propagated to the resolutions of the individual ToF stations lead to expected values of 21~$\pm$~3~ps for side A and 28~$\pm$~4~ps for side C, and to the combined $\zvtxkarel$ resolution of 5.3~$\pm$~0.6~mm.
 
The ToF detector is found to provide a good description of the time dependence of the beam-spot $z$-position in three studied (late) 2017 runs. At a more detailed level, particularly when comparing the $z$-coordinate of the primary vertex measured by the ATLAS tracker and the ToF detector on an event-by-event basis, an excess over the background level is observed in the $z_{\mbox{\tiny ATLAS}} - z_{\mbox{\tiny ToF}}$ distribution in one of these runs. This excess is present due to a contribution of central diffraction processes treated as a signal component in the subsequent statistical analysis. The width of the signal component of $6.0 \pm 2.0$~mm is in agreement, within uncertainties, with the resolution expected from single channels.
By testing the persistence of the signal against AFP track multiplicity cuts, it is concluded that the beam-induced halo activity seen by the ToF detectors constitutes a large part of the background. While the ToF detectors performed well in the other two late 2017 runs in all analyses based on single-tag events, extracting signal parameters from double-tag events with an acceptable level of confidence proved difficult. Despite the observed challenges and the low detection efficiency in 2017, it can be concluded that the ToF detector demonstrated promising precision in measuring the vertex position.

To preserve a good level of generality but also for the reason of available statistics, a limited set of cuts was used in the vertex-matching study. When more data are available, and also in physical analyses with specific final states of interest, more stringent cuts can be used, especially the matching between measurements in AFP and the central detector, which can be expected to lead to a more effective suppression of backgrounds.
 
It is worth mentioning that a measurement of $\zvtxkarel$ alone does not in itself provide a measurement of the time of an interaction in the ATLAS tracker. It would be interesting to measure the precise interaction time and use this to evaluate the time of arrival of a single diffractively emitted proton in one of the ToF detectors so as to accept only those events where there is agreement.
Such a time measurement in the central detector is planned for a future upgrade to the ATLAS detector, see Ref.~\cite{HGTD}.

\section*{Acknowledgements}


We thank CERN for the very successful operation of the LHC and its injectors, as well as the support staff at
CERN and at our institutions worldwide without whom ATLAS could not be operated efficiently.
 
The crucial computing support from all WLCG partners is acknowledged gratefully, in particular from CERN, the ATLAS Tier-1 facilities at TRIUMF/SFU (Canada), NDGF (Denmark, Norway, Sweden), CC-IN2P3 (France), KIT/GridKA (Germany), INFN-CNAF (Italy), NL-T1 (Netherlands), PIC (Spain), RAL (UK) and BNL (USA), the Tier-2 facilities worldwide and large non-WLCG resource providers. Major contributors of computing resources are listed in Ref.~\cite{ATL-SOFT-PUB-2023-001}.
 
We gratefully acknowledge the support of ANPCyT, Argentina; YerPhI, Armenia; ARC, Australia; BMWFW and FWF, Austria; ANAS, Azerbaijan; CNPq and FAPESP, Brazil; NSERC, NRC and CFI, Canada; CERN; ANID, Chile; CAS, MOST and NSFC, China; Minciencias, Colombia; MEYS CR, Czech Republic; DNRF and DNSRC, Denmark; IN2P3-CNRS and CEA-DRF/IRFU, France; SRNSFG, Georgia; BMBF, HGF and MPG, Germany; GSRI, Greece; RGC and Hong Kong SAR, China; ISF and Benoziyo Center, Israel; INFN, Italy; MEXT and JSPS, Japan; CNRST, Morocco; NWO, Netherlands; RCN, Norway; MEiN, Poland; FCT, Portugal; MNE/IFA, Romania; MESTD, Serbia; MSSR, Slovakia; ARRS and MIZ\v{S}, Slovenia; DSI/NRF, South Africa; MICINN, Spain; SRC and Wallenberg Foundation, Sweden; SERI, SNSF and Cantons of Bern and Geneva, Switzerland; MOST, Taipei; TENMAK, T\"urkiye; STFC, United Kingdom; DOE and NSF, United States of America.
 
Individual groups and members have received support from BCKDF, CANARIE, CRC and DRAC, Canada; CERN-CZ, PRIMUS 21/SCI/017 and UNCE SCI/013, Czech Republic; COST, ERC, ERDF, Horizon 2020, ICSC-NextGenerationEU and Marie Sk{\l}odowska-Curie Actions, European Union; Investissements d'Avenir Labex, Investissements d'Avenir Idex and ANR, France; DFG and AvH Foundation, Germany; Herakleitos, Thales and Aristeia programmes co-financed by EU-ESF and the Greek NSRF, Greece; BSF-NSF and MINERVA, Israel; Norwegian Financial Mechanism 2014-2021, Norway; NCN and NAWA, Poland; La Caixa Banking Foundation, CERCA Programme Generalitat de Catalunya and PROMETEO and GenT Programmes Generalitat Valenciana, Spain; G\"{o}ran Gustafssons Stiftelse, Sweden; The Royal Society and Leverhulme Trust, United Kingdom.
 
In addition, individual members wish to acknowledge support from CERN: European Organization for Nuclear Research (CERN PJAS); Chile: Agencia Nacional de Investigaci\'on y Desarrollo (FONDECYT 1190886, FONDECYT 1210400, FONDECYT 1230987); China: National Natural Science Foundation of China (NSFC - 12175119, NSFC 12275265, NSFC-12075060); European Union: European Research Council (ERC - 948254, ERC 101089007), Horizon 2020 Framework Programme (MUCCA - CHIST-ERA-19-XAI-00), Italian Center for High Performance Computing, Big Data and Quantum Computing (ICSC, NextGenerationEU); France: Agence Nationale de la Recherche (ANR-20-CE31-0013, ANR-21-CE31-0013, ANR-21-CE31-0022), Investissements d'Avenir Labex (ANR-11-LABX-0012); Germany: Baden-Württemberg Stiftung (BW Stiftung-Postdoc Eliteprogramme), Deutsche Forschungsgemeinschaft (DFG - 469666862, DFG - CR 312/5-2); Italy: Istituto Nazionale di Fisica Nucleare (ICSC, NextGenerationEU); Japan: Japan Society for the Promotion of Science (JSPS KAKENHI Grant No. 22KK0227, JSPS KAKENHI JP21H05085, JSPS KAKENHI JP22H01227, JSPS KAKENHI JP22H04944); Netherlands: Netherlands Organisation for Scientific Research (NWO Veni 2020 - VI.Veni.202.179); Norway: Research Council of Norway (RCN-314472); Poland: Polish National Agency for Academic Exchange (PPN/PPO/2020/1/00002/U/00001), Polish National Science Centre (NCN 2021/42/E/ST2/00350, NCN OPUS nr 2022/47/B/ST2/03059, NCN UMO-2019/34/E/ST2/00393, UMO-2020/37/B/ST2/01043, UMO-2021/40/C/ST2/00187, UMO-2022/47/O/ST2/00148); Slovenia: Slovenian Research Agency (ARIS grant J1-3010); Spain: BBVA Foundation (LEO22-1-603), Generalitat Valenciana (Artemisa, FEDER, IDIFEDER/2018/048), La Caixa Banking Foundation (LCF/BQ/PI20/11760025), Ministry of Science and Innovation (MCIN \& NextGenEU -PCI2022-135018-2, MICIN \& FEDER -PID2021-125273NB, RYC2019-028510-I, RYC2020-030254-I), PROMETEO and GenT Programmes Generalitat Valenciana (CIDEGENT/2019/023, CIDEGENT/2019/027); Sweden: Swedish Research Council (VR 2018-00482, VR 2022-03845, VR 2022-04683, VR grant 2021-03651), Knut and Alice Wallenberg Foundation (KAW 2017.0100, KAW 2018.0157, KAW 2018.0458, KAW 2019.0447); Switzerland: Swiss National Science Foundation (SNSF - PCEFP2\_194658); United Kingdom: Leverhulme Trust (Leverhulme Trust RPG-2020-004); United States of America: Neubauer Family Foundation.


\clearpage
\appendix

\printbibliography

\clearpage
\input{atlas_authlist}

\end{document}

%% file: atlas_authlist.tex
 
\begin{flushleft}
\hypersetup{urlcolor=black}
{\Large The ATLAS Collaboration}

\bigskip

\AtlasOrcid[0000-0002-6665-4934]{G.~Aad}$^\textrm{\scriptsize 102}$,
\AtlasOrcid[0000-0001-7616-1554]{E.~Aakvaag}$^\textrm{\scriptsize 16}$,
\AtlasOrcid[0000-0002-5888-2734]{B.~Abbott}$^\textrm{\scriptsize 120}$,
\AtlasOrcid[0000-0002-1002-1652]{K.~Abeling}$^\textrm{\scriptsize 55}$,
\AtlasOrcid[0000-0001-5763-2760]{N.J.~Abicht}$^\textrm{\scriptsize 49}$,
\AtlasOrcid[0000-0002-8496-9294]{S.H.~Abidi}$^\textrm{\scriptsize 29}$,
\AtlasOrcid[0000-0002-9987-2292]{A.~Aboulhorma}$^\textrm{\scriptsize 35e}$,
\AtlasOrcid[0000-0001-5329-6640]{H.~Abramowicz}$^\textrm{\scriptsize 151}$,
\AtlasOrcid[0000-0002-1599-2896]{H.~Abreu}$^\textrm{\scriptsize 150}$,
\AtlasOrcid[0000-0003-0403-3697]{Y.~Abulaiti}$^\textrm{\scriptsize 117}$
% ...
\bigskip
\\

% ...
$^{16}$Department for Physics and Technology, University of Bergen, Bergen; Norway.\\
% ...
$^{29}$Physics Department, Brookhaven National Laboratory, Upton NY; United States of America.\\
% ...
$^{35}$$^{(a)}$Facult\'e des Sciences Ain Chock, R\'eseau Universitaire de Physique des Hautes Energies - Universit\'e Hassan II, Casablanca;$^{(b)}$Facult\'{e} des Sciences, Universit\'{e} Ibn-Tofail, K\'{e}nitra;$^{(c)}$Facult\'e des Sciences Semlalia, Universit\'e Cadi Ayyad, LPHEA-Marrakech;$^{(d)}$LPMR, Facult\'e des Sciences, Universit\'e Mohamed Premier, Oujda;$^{(e)}$Facult\'e des sciences, Universit\'e Mohammed V, Rabat;$^{(f)}$Institute of Applied Physics, Mohammed VI Polytechnic University, Ben Guerir; Morocco.\\
% ...
$^{49}$Fakult\"{a}t Physik , Technische Universit{\"a}t Dortmund, Dortmund; Germany.\\
% ...
$^{55}$II. Physikalisches Institut, Georg-August-Universit\"{a}t G\"ottingen, G\"ottingen; Germany.\\
% ...
$^{102}$CPPM, Aix-Marseille Universit\'e, CNRS/IN2P3, Marseille; France.\\
% ...
$^{117}$Department of Physics, New York University, New York NY; United States of America.\\
% ...
$^{120}$Homer L. Dodge Department of Physics and Astronomy, University of Oklahoma, Norman OK; United States of America.\\
% ...
$^{150}$Department of Physics, Technion, Israel Institute of Technology, Haifa; Israel.\\
$^{151}$Raymond and Beverly Sackler School of Physics and Astronomy, Tel Aviv University, Tel Aviv; Israel.\\
% ...
$^{a}$ Also Affiliated with an institute covered by a cooperation agreement with CERN.\\
$^{b}$ Also at An-Najah National University, Nablus; Palestine.\\
$^{c}$ Also at Borough of Manhattan Community College, City University of New York, New York NY; United States of America.\\
$^{d}$ Also at Center for High Energy Physics, Peking University; China.\\
$^{e}$ Also at Center for Interdisciplinary Research and Innovation (CIRI-AUTH), Thessaloniki; Greece.\\
$^{f}$ Also at Centro Studi e Ricerche Enrico Fermi; Italy.\\
% ...

\end{flushleft}
